\documentclass[aps,superscriptaddress,twocolumn,floatfix,citeautoscript,longbibliography,hyperlinks]{revtex4-2}
\usepackage{graphicx}
\usepackage{dcolumn}
\usepackage[dvipsnames]{xcolor}
\usepackage[colorlinks,citecolor=blue,urlcolor=blue]{hyperref}
\usepackage{amssymb,amsmath,bm,bbm,mleftright,mathtools}
\DeclareMathAlphabet\mathbfcal{OMS}{cmsy}{b}{n}
\newcommand{\bra}[1]{\left\langle #1\right|}
\newcommand{\ket}[1]{\left|#1\right\rangle}

\renewcommand{\vec}{\mathbf}
\makeatletter
\newsavebox{\@brx}
\newcommand{\llangle}[1][]{\savebox{\@brx}{\(\m@th{#1\langle}\)}%
  \mathopen{\copy\@brx\kern-0.5\wd\@brx\usebox{\@brx}}}
\newcommand{\rrangle}[1][]{\savebox{\@brx}{\(\m@th{#1\rangle}\)}%
  \mathclose{\copy\@brx\kern-0.5\wd\@brx\usebox{\@brx}}}
\makeatother

\begin{document}

\title{Postselection-free approach to monitored quantum dynamics and entanglement phase transitions}

\author{ Kim P\"oyh\"onen}
\affiliation{Computational Physics Laboratory, Physics Unit, Faculty of Engineering and
Natural Sciences, Tampere University, FI-33014 Tampere, Finland}
\affiliation{Helsinki Institute of Physics, FI-00014 University of Helsinki, Finland}

\author{Ali G. Moghaddam}\email{Email: a.ghorbanzade@gmail.com}
\affiliation{Computational Physics Laboratory, Physics Unit, Faculty of Engineering and
Natural Sciences, Tampere University, FI-33014 Tampere, Finland}
\affiliation{Helsinki Institute of Physics, FI-00014 University of Helsinki, Finland}
\address{Department of Applied Physics, Aalto University, P.O. Box 11000, FI-00076 Aalto, Finland}

\author{Moein N. Ivaki}
\affiliation{Department of Applied Physics, Aalto University, P.O. Box 11000, FI-00076 Aalto, Finland}
\affiliation{Quantum Technology Finland Center of Excellence, Department of Applied Physics, Aalto University,
P.O. Box 11000, FI-00076 Aalto, Finland}

\author{Teemu Ojanen} \email{Email: teemu.ojanen@tuni.fi}
\affiliation{Computational Physics Laboratory, Physics Unit, Faculty of Engineering and
Natural Sciences, Tampere University, FI-33014 Tampere, Finland}
\affiliation{Helsinki Institute of Physics, FI-00014 University of Helsinki, Finland}

\begin{abstract}
Measurement-induced entanglement phase transitions in monitored quantum circuits have stimulated activity in a diverse research community. However, the study of measurement-induced dynamics, due to the requirement of exponentially complex postselection, has been experimentally limited to small or specially designed systems that can be efficiently simulated classically. We present a solution to this outstanding problem by introducing a scalable protocol in $U(1)$ symmetric circuits that facilitates the observation of entanglement phase transitions \emph{directly} from experimental data, without detailed assumptions of the underlying model or benchmarking with simulated data. Thus, the method is applicable to circuits which do not admit efficient classical simulation and allows a reconstruction of the full entanglement entropy curve with minimal theoretical input. Our approach relies on adaptive circuits and a steering protocol to approximate pure-state trajectories with mixed ensembles, from which one can efficiently filter out the subsystem $U(1)$ charge fluctuations of the target trajectory to obtain its entanglement entropy. The steering protocol replaces the exponential costs of postselection and state tomography with a scalable overhead which, for fixed accuracy $\epsilon$ and circuit size $L$, scales as $\mathcal{N}_s\sim L^{5/2}/\epsilon$.
\end{abstract}
\maketitle

\section{introduction}
Quantum dynamics and phase transitions in monitored quantum circuits have rapidly stimulated an avalanche of interest, bringing together researchers of quantum information, statistical physics and condensed matter physics~\cite{skinner_measurement-induced_2019,li_measurement-driven_2019,Fisher_random_quantum_circuits,potter_entanglement_2021,zabalo_critical_2020,gullans_dynamical_2020,Schomerus_weak_measurements_2019,lavasani_measurement-induced_2021,Turkeshi2021zeroclick}. The topic bridges multiple fields, providing a treasure trove of new phenomena and concepts. Moreover, these developments have concretely demonstrated that some of the most fascinating realizations of quantum matter are presently provided by the emerging Noisy Intermediate-scale Quantum (NISQ) devices~\cite{preskill2018NISQ, RevModPhys.94.015004}.

The interest towards monitored circuits exploded after the discovery of measurement-induced entanglement phase transitions~\cite{li_quantum_2018,cao_entanglement_2019,PhysRevB.100.064204, PhysRevX.11.011030,chan_unitary-projective_2019,Jian-Ludwig2020_measurement_criticality,Ashida2020PRB,bao_theory_2020,choi_quantum_2020,Mirlin2023PRX}. The proliferation of entanglement due to local entangling operations is first hindered, and then completely halted, by measurement of a fraction of the qubits in the system. This manifests in a phase transition between the entanglement entropy volume-law and area-law phases when the measured fraction is increased. However, direct experimental approaches to entanglement phase transitions face two fundamental obstacles. First, a direct measurement of entanglement entropy requires quantum state tomography, the cost of which scales exponentially with system size, an issue already faced in some recent experiments~\cite{Monroe_measurement_2022,google2023measurement,Minnich_2023measurement}. The second, and more serious, bottleneck is posed by the exponential postselection problem~\cite{Fisher_random_quantum_circuits}: to extract the properties of a quantum state resulting from monitored dynamics, one needs to prepare an ensemble of states with precisely the same measurement outcomes throughout the temporal evolution. Even for modest systems of 10-20 qubits, the cost of preparing a postselected ensemble soon becomes astronomical. 
\begin{figure}[t]
    \centering
    \includegraphics[width=.99\columnwidth]{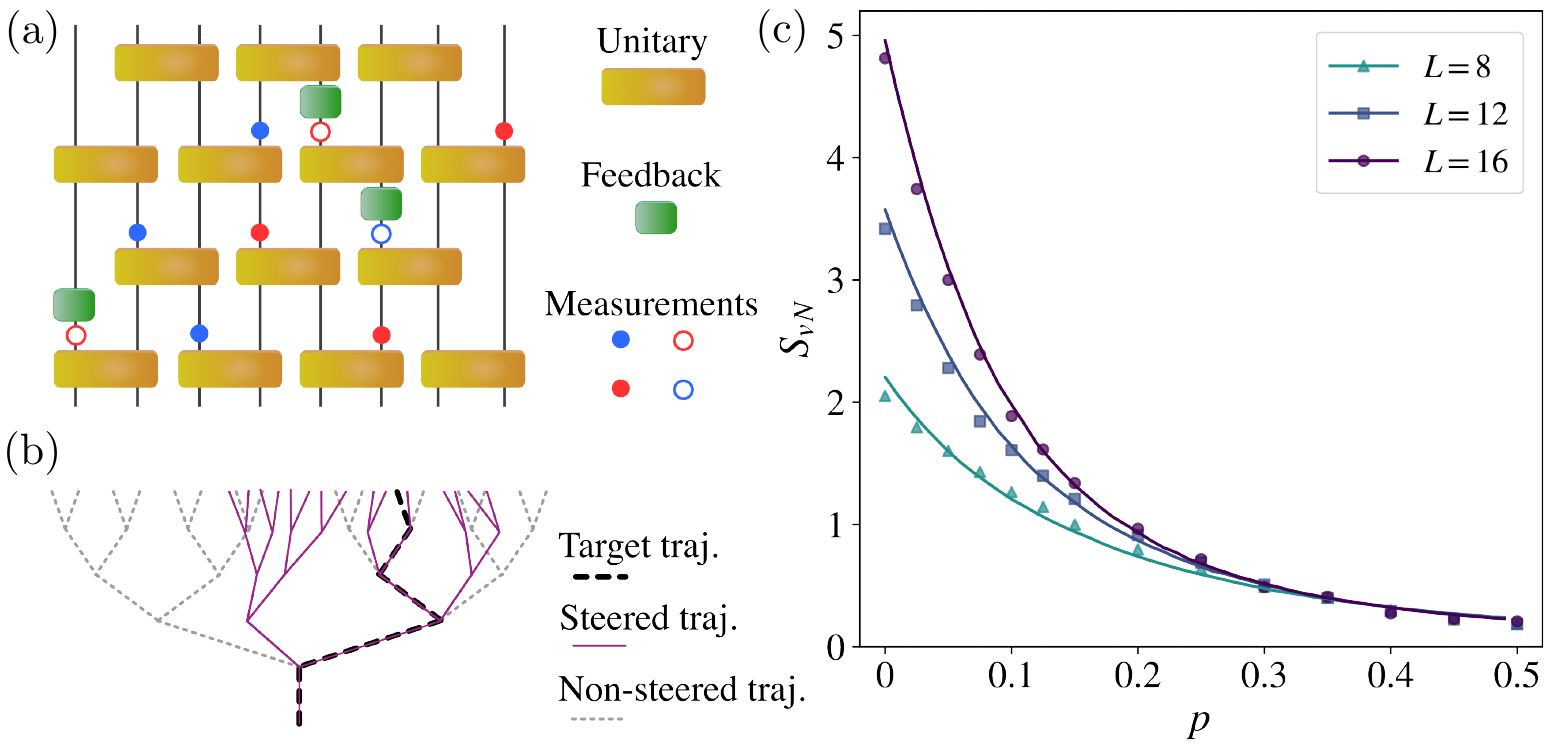}  
    \caption{\textbf{Scalable quantum simulation of entanglement in monitored quantum circuits with $U(1)$ symmetry.} \textbf{a,} The studied brickwork circuits with two-qubit random unitaries and projective measurements. To implement steering towards a chosen target state, conditional single-qubit feedforward gates are applied to locally correct the state after measurements to match those of the target trajectory.
    \textbf{b,} Schematic illustration of branching of quantum trajectories by measurements, and steering them towards the target trajectory. 
    \textbf{c,} Comparison of the exact trajectory-averaged entanglement entropy (solid lines) and the output of the protocol developed in this work combined with the known flucutation-entropy relation (markers).}
    \label{fig1}
\end{figure}
The existing workarounds to experimentally probe the transition rely on relaxing the postselection condition and pinpointing the transition by cross-referencing the measurement data with classical simulation data and theoretical mappings~\cite{kamakari2024experimental_scalable,PhysRevLett.130.220404, gullans_scalable_2020,Ippoliti_postselection, Gullans2023neural,Garratt2023,PRXQuantum.5.020347,buchhold2022revealing,Sierant_Turkeshi_2023}. They can be regarded as hybrid experimental-theoretical approaches which rely heavily on theoretical assumptions on the underlying experimental systems and typically require that the systems can be efficiently simulated classically, involve heavy post-processing and provide only partial information of the entanglement dynamics. 

In this work we introduce a scalable quantum simulation approach to reconstructing the entanglement entropy in $U(1)$-symmetric monitored circuits \emph{without exponential cost} from either postselection or quantum state tomography. Our approach, illustrated in Fig.~\ref{fig1}, employs adaptive circuits to remove the need for exponential postselection. We show how the relevant properties of a single quantum trajectory can be efficiently filtered from experimental data. Thus, our work establishes a scalable method to observing the measurement-induced entanglement phase transitions \emph{directly} from the experimental data, \emph{without} requiring classical benchmarking or detailed assumptions of the experimental system. With minimal theoretical input, our procedure also allows a full reconstruction of the entanglement entropy curve for all measurement rates.

\section{Entanglement dynamics from $U(1)$ charge fluctuations} \label{sec:II}

\subsection{Full postselection}\label{subsec:postsel}

To set the stage, we briefly review how the  entanglement entropy can be extracted from the $U(1)$ charge fluctuations in  $U(1)$ symmetric circuits as discussed in Refs.~\cite{PRL2023,Oshima2023,oulu_feedback2024}. The studied model consist of a linear array of $L$ qubits. At each time step, the system is evolved by applying unitary two-qubit gates followed by projective single-qubit measurements. The two-qubit gates, at even and odd time steps, act on qubits connected by even and odd links, as seen in Fig. \ref{fig1}. 
The successive odd and even time steps constitute a full cycle $t$. Without measurement, the evolution of a cycle is generated by  
\begin{equation} \label{eq:evolution}
    {\bf U}(t) =\prod_{\text{even } n}U_{n,n+1}(2t) \prod_{\text{odd } n}U_{n,n+1}(2t-1)
\end{equation}
where $U_{n,n+1}(\tau)$ is a two-qubit unitary acting on qubits at positions $n$ and $n+1$ at time step $\tau=2t\:\rm{or}\:2t-1$. To implement the $U(1)$ charge $Z_{L}= \sum_{n} Z_{n}$ conservation, where $Z_n$ is the Pauli-Z matrix operating on the $n$'th qubit, these unitary gates take the form
\begin{equation}
    U_{n,n+1} = \begin{pmatrix}
    e^{i\varphi_{00}} & &\\
     &  e^{i\varphi_{11}} &\\
     && {\cal U}_{2\times 2}
    \end{pmatrix}
\end{equation}
in the basis $\{\ket{00}, \ket{11}, \ket{01}$, $\ket{10}\}$. Here ${\cal U}_{2\times 2}$ is a generic $2\times 2$ unitary matrix constituting of four independent phases. Each local unitary gate $U_{n,n+1}(\tau)$ is chosen randomly and independently, by sampling all six phases that parameterize it from a uniform random distribution. Then, single-qubit measurements occur at the end of each half-cycle over randomly-chosen qubits with probability $p$.
These measurements make the full dynamics non-unitary and inherently probabilistic. Considering single-qubit projective measurements, the wave function randomly collapses to an eigenstate of the corresponding single-qubit observable
    $\ket\Psi \to \frac{1}{\lVert  P_{n,\alpha}\ket\Psi \rVert } P_{n,\alpha}\ket\Psi$
with projectors $P_{n,\alpha}$ acting on the qubit $n$ and satisfying normalization $\sum_{\alpha} P_{n,\alpha} = {\mathbb I} $. We assume all measurements are performed in the $Z$-basis with the projectors $P_{n,\pm} = \big({\mathbb I} \pm Z_{n}\big)/2$.
Each set of particular measurement outcomes ${\bf m}\in \{1,-1\}^M$, where $M$ qubits are measured during the evolution, defines a unique quantum trajectory throughout which the state remains pure. The evolved state for any single realization of the monitored circuit can be denoted by $\ket{\Psi}_{\bf U,m}$ for a given random set of unitaries ${\bf U}$ and measurement outcomes ${\bf m}$.
\begin{figure}[ht]
    \centering
    \includegraphics[width=.95\columnwidth]{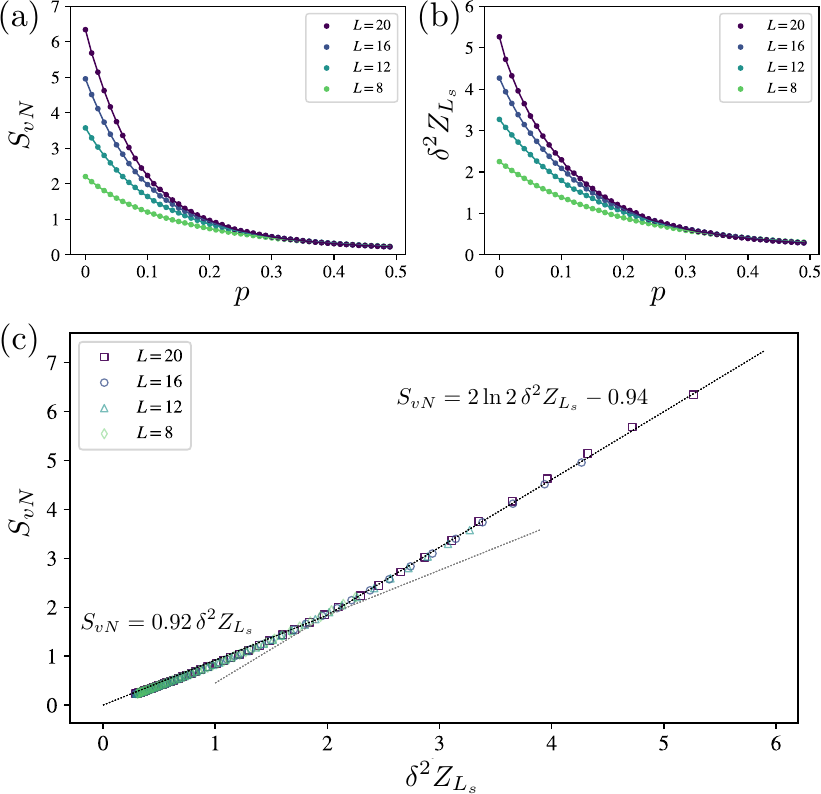}  
    \caption{\textbf{Pure state fluctuation-entanglement correspondence.} \textbf{a,} Trajectory-averaged entanglement entropy as a function of the single-qubit measurement rate. \textbf{b,} Same as \textbf{a} but for the subsystem charge fluctuations. \textbf{c,} Universal variance-entropy relation using the data from figs. \textbf{a}-\textbf{b}, which is excellently approximated by two linear regimes. }
    \label{fig:fluc1}
\end{figure}

To study the entanglement dynamics, we choose the initial state $\ket{\Psi} = \ket{\uparrow\downarrow\uparrow\downarrow \cdots \uparrow\downarrow}$, which, due to conservation, fixes the total $U(1)$ charge to zero $Z_{L}= \sum_{n} Z_{n}=0$. The initial state is then propagated forward until observables averaged over different trajectories saturate to a stationary value. In practice, this happens after $\sim L$ cycles. More details on the numerical calculations are presented in App.~\ref{app_A}. The von Neumann entanglement entropy for a subsystem with length $L_s$ is $S_\text{vN} = - {\rm Tr}\big(\rho_{s}\ln \rho_s\big)$, where the reduced density matrix $\rho_s$ is obtained from a pure state quantum trajectory by tracing out the complement as $\rho_s = {\rm Tr}_{\bar{s}}(\ket{\Psi}\bra{\Psi})$. Below we exclusively consider partitions which divide the system in half $L_s=L/2$. As shown in Ref.~\cite{PRL2023}, the entanglement entropy exhibits a phase transition between a volume-law and an area-law phase at $p=p_c\approx 0.14 $ with the critical exponent $\nu\approx1.3$. However, directly measuring the entanglement entropy of a quantum state is challenging, requiring tomography methods that scale poorly with system size. For system with $U(1)$ charge conservation, this is simplified by the fact that the fluctuations $\delta^2 Z_{L_s}=\langle Z_{L_s}^2 \rangle - \langle Z_{L_s}\rangle^2$ of the subsystem charge $Z_{L_s}= \sum_{n\in L_s} Z_{n}$ exhibit the same scaling behavior as the entanglement entropy, as is illustrated in Fig.~\ref{fig:fluc1}a-b. While a single-parameter scaling analysis (as discussed in App.~\ref{app:B}) cannot provide conclusive evidence of transitions for the available system sizes, in App.~\ref{app:D} we provide a general argument that the bipartite charge fluctuations and the entanglement entropy exhibit the same phase diagram and $p_c$.  As seen in Fig.~\ref{fig:fluc1}c and discussed in App.~\ref{app:c}, the entanglement entropy is a system size- and $p$-independent universal function of the variance $S_\text{vN}=f\left(\delta^2 Z_{L_s}\right)$ which, to an excellent approximation, is piecewise linear  $S_\text{vN}= a\,\delta^2 Z_{L_s} $ for $0\leq \delta^2 Z_{L_s}\leq 2 $ and  $S_\text{vN}= 2\ln 2\, \delta^2 Z_{L_s} +(2a-4\ln 2)$ for $ \delta^2 Z_{L_s}> 2 $ with $a=0.92$. The fact that the entanglement vs. charge fluctuation data in Fig.~\ref{fig:fluc1}c collapses on a single curve \emph{without any fitting parameters} requires that the two quantities are linearly related $S_\text{vN}\sim 2\ln 2\, \delta^2 Z_{L_s}$ for large values. This linear relation necessitates that the transition for both quantities belong to the same universality class. As discussed in App.~\ref{app:c}, the same observation also holds for the 2nd R\'enyi entropy. The parameter-free collapse in Fig.~\ref{fig:fluc1}c provides strong evidence, supporting the general argument in App.~\ref{app:D} that the relevant properties of the entanglement phase transition can be obtained from bipartite charge fluctuations. Additionally, the variance-entropy relation allows a simple reconstruction of the full entanglement curve when $ \delta^2 Z_{L_s}$ is known. However, unlike the phase transition, obtaining the variance-entropy coefficients does require simulation or post-selection at least for small system sizes.

It should be emphasized that the discussion above exclusively concerns \emph{bipartite} charge fluctuations and associated volume-area law transitions within \emph{pure states} of a fixed charge sector. This phenomenon is fundamentally distinct from the charge-sharpening transition \cite{Agrawal_2022}, which involves the \emph{total} charge dynamics of mixed states (or superpositions across sectors) and acts as a precursor to purification, despite thematic similarities due to importance of $U(1)$ conservation. Strong arguments, detailed in App. ~\ref{app:D}, indicate that the bipartite charge fluctuation transitions studied here coincide with entanglement entropy transitions, where entanglement area/volume laws in random quantum circuits imply corresponding behaviors for the subsystem charge fluctuation as well. This expected coincidence further highlights the distinction from charge sharpening, as the latter precedes the entanglement transition \cite{ippoliti2023learnability,PhysRevLett.129.200602}. Similarly, we emphasize that the mixed-state ensemble introduced below serves solely as a technical tool to accurately reproduce the here discussed pure-state bipartite charge fluctuations, and should not be confused with an ensemble used to study \emph{total} charge dynamics and the associated charge-sharpening transition.


\subsection{Relaxing the exponential postselection by steering}\label{subsec:steering}

\begin{figure}
    \centering
    \includegraphics[width=.95\columnwidth]{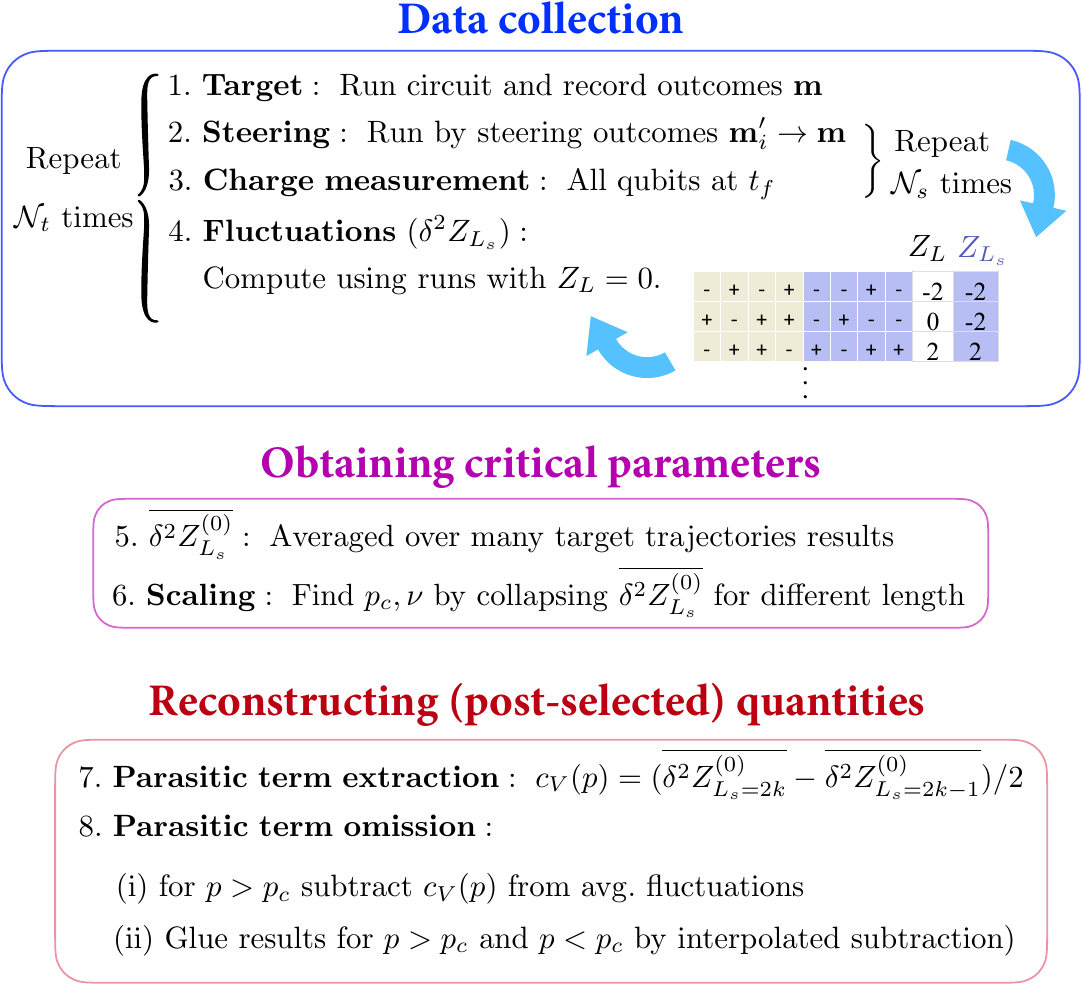}
    \caption{\textbf{Summary of the scalable steering protocol.} It provides the recipe to observe entanglement phase transitions and reconstruct the entanglement entropy in monitored circuits. }
    \label{fig:recipe}
\end{figure}

As seen above, the fluctuations of the $U(1)$ charge allows one to access the entanglement entropy of pure states without exponential complexity associated with the state tomography. Nevertheless, one still has to deal with the postselection problem over quantum trajectories $\ket{\Psi}_{\bf U, m}$ corresponding to specific unitaries and $2^M$ different possible measurement outcomes ${\bf m}$. To enable fully scalable experimental studies of measurement-induced dynamics, we need a procedure to simulate properties of a single target quantum trajectory $\ket{\Psi_{\bf U, m}}$ without the need for exponential postselection. 
Instead of running the circuit exponentially many times in the hope of reproducing the target trajectory multiple times, we adopt the adaptive steering protocol depicted in Fig.~\ref{fig1}b and summarized in Fig.~\ref{fig:recipe}. During the circuit execution, whenever the obtained measurement outcome differs from the one in the target trajectory, the circuit performs a Pauli-X operation on the measured qubit. This operation locally steers the outcomes immediately after measurements to match those of the \emph{target trajectory}, ${\bf m}$. Repeating the steering evolution $\mathcal{N}_s$ times, each time we get a possibly different state $\ket{\Psi_{{\bf m}^\prime_i\to{\bf m}}}$ where ${\bf m}^\prime_i$ denotes the measurement outcomes in the $i$'th run. Without any postselection, the repetition of steering runs results in a mixed ensemble $\rho_{\bf U,m}=\frac{1}{\mathcal{N}_s}\sum_i 
\ket{\Psi_{{\bf m}^\prime_i\to{\bf m}}}\bra{\Psi_{{\bf m}^\prime_i\to{\bf m}}}
$ associated with the target state $\ket{\Psi_{\bf U,m}}$ \footnote{Note that the index $U$ for steered states has been dropped for brevity}.
Steering processes by Pauli-X operators break the $U(1)$ charge conservation, creating an incoherent mixture of charge sectors during the evolution. However, the density matrix still commutes with the total charge, $[\rho_{\bf U,m}, Z_L] = 0$, and can be block-diagonalized in the total charge sectors. This allows for the separation of the steered states according to their total charge $Z_L$, resulting in steered ensembles, $\tilde{\rho}_{\bf U,m}^{Z_L}=\frac{1}{\mathcal{N}_{Z_L}}\sum^\prime_{i} 
\ket{\Psi_{{\bf m}^\prime_i\to{\bf m}}}\bra{\Psi_{{\bf m}^\prime_i\to{\bf m}}}
$ by summing over the states having a fixed total charge. We see that this way we can generate a mixed state which approximates some \emph{statistical properties} of specific target states. This approach is different from the study of a mixed steered ensemble itself; the mixed state is introduced only for the purpose of obtaining the charge fluctuations in the target state, and its other properties are not studied. This is to be compared with the post-selection method, where a large number of identical states is created and measured instead.

\begin{figure*}[ht]
    \centering
    \includegraphics[width=1.96\columnwidth]{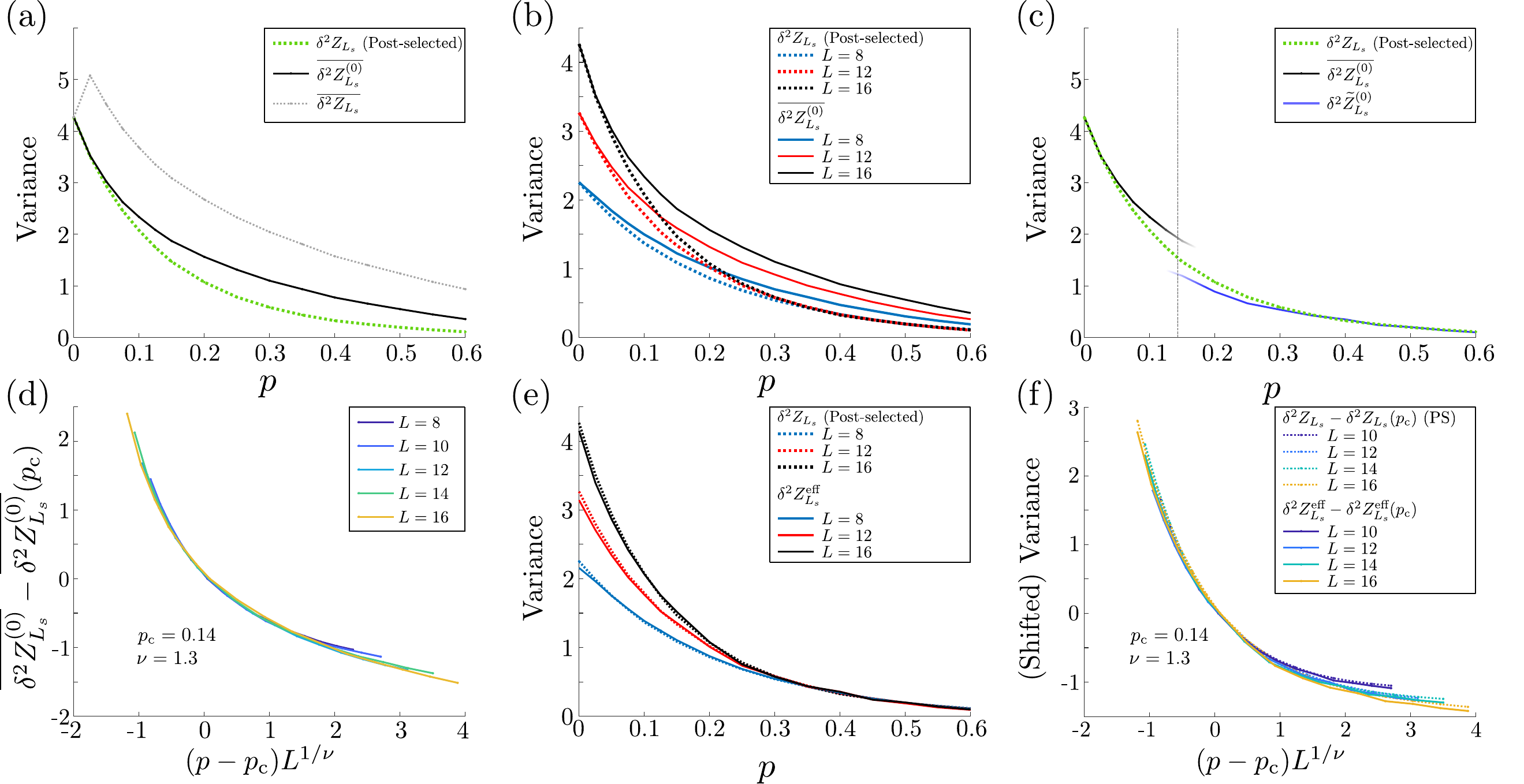}  
\caption{\textbf{Extracting postselected quantities and entanglement phase transitions from steered ensembles.} \textbf{a,} Subsystem charge fluctuations as a function of the measurement rate for $15 \times 1000$ target trajectories (see App.~\ref{app_A} for details). Comparison of postselected data (dotted line) with the value obtained from the full steered ensemble $\overline{\delta^2 Z_{L_s}}$ and the $Z_L=0$ sector value $\overline{\delta^2 Z_{L_s}^{(0)}}$ for system size $L=16$. \textbf{b,} Comparison of the postselected and the $Z_L=0$ sector fluctuations for system lengths $8, 12, 16$, illustrating the parasitic volume contribution in the regime which for postselected data obeys the area law. \textbf{c,} Postselected data at $L=16$ compared to both $\delta^2 Z_{L_s}^{Z_L = 0}$, which is a good approximation for low $p$, and to the area-law corrected term $\delta^2 Z_{L_s}^{Z_L = 0} - c_V(p)$ which is a good approximation at high $p$. Around $p_c$, shown as a vertical line, they each deviate from the postselected value. \textbf{d,} Scaling collapse of the $Z_L=0$ sector fluctuations, which exhibit criticality with the same $p_\mathrm{c}$ and $\nu$ as the postselected fluctuations. \textbf{e,} Comparison of postselected fluctuations (dotted lines) and the effective fluctuations (solid lines) obtained through Eq.~\eqref{eq:glue} for system lengths $L = 8,12,16$. For the clarity, odd subsystem lengths have been excluded; the match is equally good for those. \textbf{f,} Scaling collapse of the effective fluctuation data from \textbf{e}, compared to the scaling collapse curves for the postselected data.}
    \label{fig4}
\end{figure*}

While the reduced von Neumann entropy is no longer an appropriate measure of entanglement for a mixed state, we now illustrate how to extract the postselected charge fluctuations from the fluctuations of the steered ensemble $\tilde{\rho}_{\bf U,m}^{Z_L}$. Since the postselected fluctuations correspond linearly to the entanglement entropy, as shown in Fig.~\ref{fig:fluc1}c, obtaining a good estimate of the post-selected fluctuations from the mixed state provides access to the entanglement phase transition. The main issue which needs to be resolved is that the fluctuations of the steered ensemble contain an additional incoherent contribution which is expected to lead to parasitic volume-law fluctuations. As discussed in App.~\ref{app:E}, this parasitic contribution is smaller for the charge-sector filtered ensembles $\tilde{\rho}^{Z_L}$ compared to the full steering mixture $\rho$ (we have dropped the subscripts for simplicity). Specifically, as shown in Fig.~\ref{fig4}a, the fluctuations corresponding to $\tilde{\rho}^{Z_L=0}$ provide an excellent approximation of the postselected fluctuations in the volume-law regime. In the area-law regime, however, the incoherent contribution has to be explicitly subtracted. As discussed in App.~\ref{app:F}, the incoherent fluctuations can be easily distinguished from the area-law contribution due to their different system-size dependency.

Steps 1.-4. of the protocol in Fig.~\ref{fig:recipe} are illustrated in Figs.~\ref{fig4}a-b. The trajectory-averaged steered ensemble fluctuations $\overline{\delta^2 Z_{L_s}^{(0)}}$ corresponding to $\tilde{\rho}^{Z_L=0}$ show an excellent match with the postselected value at small $p$, while the incoherent contribution leads to simple size-dependent overestimation in the area-law regime. The subtraction of the parasitic volume law coefficient, outlined in steps 7. and 8., is illustrated in Fig.~\ref{fig4}c; subtracting the volume-law contribution $\delta^2 \widetilde{Z}_{L_s}^{(0)}=\overline{\delta^2 Z_{L_s}^{(0)}}-c_V(p)\left[L_s-2\right]$, with the length-independent parasitic volume-law coefficient
\begin{equation}
    c_V(p) \equiv  \left(\overline{\delta^2 Z_{L_s=2k}^{(0)}}-\overline{\delta^2 Z_{L_s=2k-1}^{(0)}}\right)/2\label{eq:cvp}
\end{equation} 
for arbitrary integer $k\geq 1$, produces an excellent approximation of the postselected fluctuations in the area-law regime, in turn. Now since postselected fluctuations can be straightforwardly obtained from $\overline{\delta^2 Z_{L_s}^{(0)}}$, it can be expected to exhibit critical behavior at the same $p_c$ as the postselected fluctuations and, hence, the entanglement entropy. Indeed, as highlighted in steps 5.-6. and illustrated in Fig.~\ref{fig4}d, the transition can be observed by collapsing $\overline{\delta^2 Z_{L_s}^{(0)}}$ with the single-parameter scaling Ansatz $F(p,L)-F(p_c,L)=\tilde{F}\left[(p-p_c)L^{1/\nu}\right]$, enabling extraction of the critical rate $p_c$ and the critical exponent $\nu$ \emph{directly from experimentally obtainable data}. In practice, one can carry out the scaling analysis before reconstructing the postselected quantities, as indicated in Fig.~\ref{fig:recipe}. Finally, as stated in step 8.(ii), the separate approximations for the postselected charge fluctuations in the volume-law and area-law phases can be combined into an effective charge fluctuation
\begin{equation}
\delta^2 {Z}_{L_s}^{\rm{eff}}=\overline{\delta^2 Z_{L_s}^{(0)}}-g(p)\,c_V(p)[L_s-2],\label{eq:glue}
\end{equation}
where $g(p)$ is a smooth step-like function interpolating between 0  (for $p\ll p_c$) and 1  (for $p\gg p_c$).  As the volume-law and area-law asymptotes separately exhibit the same scaling around $p_c$, it is natural to expect that the the transition from one functional form to other is also controlled by the scaling variable. Thus, the function $g(p)$ should be of the form $g(p)=G[(p-p_c)L^{1/\nu}]$, where $G(x)$ is a smooth unit-step function centered at $x=0$. This determines the width of the transition, which approaches zero in the thermodynamic limit, and fixes $g(p)$ apart from tiny deformations. One of the most obvious candidates, $G(x)=(\tanh{x}+1)/2$, leads to the $\delta^2 {Z}_{L_s}^{\rm{eff}}$ in Fig.~\ref{fig4}e-f, which very accurately follows the postselected fluctuations in the whole $p$ range. With this result, an accurate reconstruction of the whole entanglement entropy curve can be straightforwardly obtained by employing the simple relationship between postselected fluctuations and entropy, as depicted in Fig.~\ref{fig1}c. As discussed in App.~\ref{app:g}, the steering protocol replaces the exponential complexity associated with both the postselection and the entropy measurement with a scalable overhead $\mathcal{N}_s\sim\frac{L^{5/2}}{\epsilon}$ per trajectory, where $\epsilon$ is the the desired accuracy and $L$ is the circuit size. This estimate contains the postselection cost to the zero total charge sector, which increases very mildly $\sim \sqrt{L}$ as a function of system size.

Finally, we note that a breakdown of the correspondence between the entanglement phase transition and the charge fluctuation transition in the steered ensemble would require either that 1) the correspondence between the entanglement phase transition and charge fluctuation transition breaks down in the postselected case or 2) the correspondence between the steered fluctuations and the postselected fluctuations breaks down. The general argument for coinciding phase diagrams of entropy and charge fluctuations established in App.~\ref{app:D}, together with the parameter-free entropy-fluctuation collapse summarized in Fig.~\ref{fig:fluc1}(c), provide strong evidence against 1). While our data does not lend any support for scenario 2), as seen in Fig.~\ref{fig4}(e), it is naturally impossible to rule it out for system sizes far beyond numerical approaches. However, to put this possibility in proper perspective, it is also impossible to definitely confirm the existence of the entanglement phase transition itself beyond rare tailored models. With these caveats, we conclude that the steering protocol provides a scalable method to study the entanglement phase transitions and entanglement dynamics in $U(1)$ symmetric circuits, offering new possibilities to experimentally probe the physics of monitored circuits.

\section{Discussion and outlook}
In this work, we introduced a scalable approach to the measurement-induced entanglement dynamics and entanglement phase transitions in $U(1)$-symmetric circuits. The key idea is that adaptive circuits and charge fluctuations can be employed to simultaneously avoid the exponential complexity associated both with the postselection and with quantum state tomography. From a technological point of view, adaptive circuits impose additional requirements on the NISQ devices. Adaptive dynamics are currently most effectively implemented in ion-trap simulators~\cite{Monroe_measurement_2022, experiment2023_feedback_ion_trap} and, more recently, in superconducting quantum circuits~\cite{devoret_2023_feedback}. There is also an ongoing effort to achieve corrective operations based on more complex multi-qubit measurements, which are the backbone of fully-fledged quantum error correction~\cite{google2023QEC,Terhal2015review}. The simpler adaptive functionality discussed here is a prerequisite to error correction schemes and thus actively pursued in all platforms. Furthermore, the adaptive functionality is crucial to realizing measurement-induced phases of matter and absorbing-state transitions~\cite{PRXQuantum.4.040309,Entanglement_steering,Turkeshi2023,Khemani2024,iadecola2023}.

Our approach establishes a protocol for probing measurement-induced entanglement phase transitions directly from experimental data without requiring postselection and also any additional theoretical benchmarks. This is in contrast to previous hybrid experimental-theoretical methods \cite{kamakari2024experimental_scalable,PhysRevLett.130.220404,Gullans2023neural}, and thus our protocol for obtaining the subsystem charge fluctuations is not limited to systems that admit efficient classical simulation. A more detailed description of the entropy beyond transition points does require simulation or post-selection of small systems to obtain the entropy-fluctuation coefficients as in Fig.~\ref{fig:fluc1}\textbf{c}, which can then be used to extrapolate to larger systems. While we acknowledge the current lack of full rigor in theoretically linking charge fluctuations obtained from a steered ensemble to the entanglement transition, the protocol steps are well-justified, and their plausibility is strongly supported by numerical results showing the expected volume- and area-law scaling. Furthermore, our method allows straightforward reconstruction of the full entanglement entropy curve as a function of measurement rate. Aside from the entanglement phase transition, symmetric circuits support further intriguing measurement-induced phenomena \cite{Agrawal_2022}. The generalization of the steering approach to these phenomena will be studied in future works.


\section{Acknowledgements} 
A.G.M. and T.O. acknowledge Jane and Aatos Erkko Foundation for financial support. T.O. also acknowledges the Finnish Research Council project 331094. The authors thank P. Sierant for discussions.

\appendix

\section{Note on computational details} \label{app_A}

\begin{figure}[ht]
    \centering
    \includegraphics[width=0.95\columnwidth]{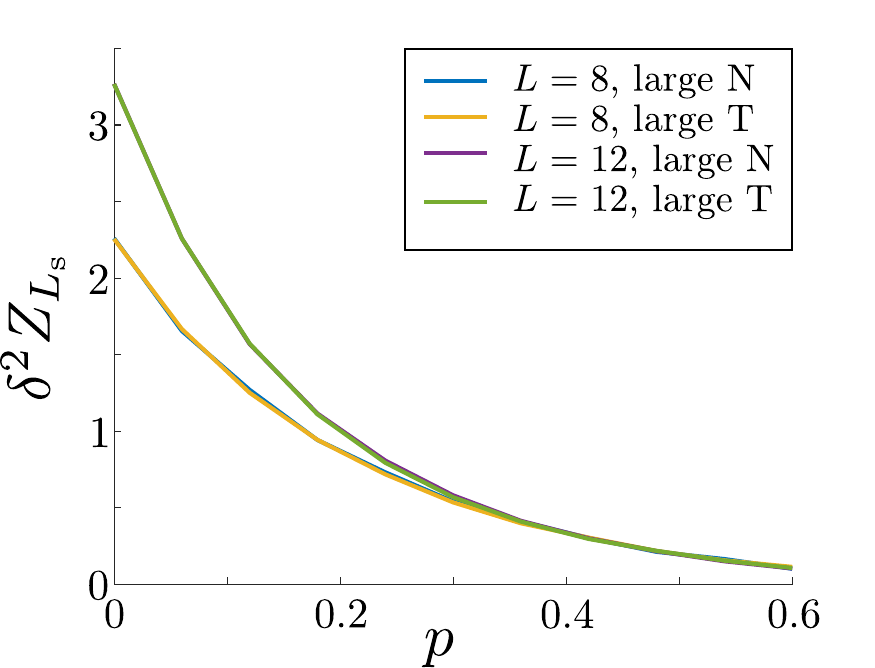}  
    \caption{\textbf{Comparison of trajectory averaging and time averaging for postselected data.} The average of the expected variance at the final timestep of 15000 different runs is contrasted with the average of the same at the end of each cycle for a single run over 15000 cycles. 
    \label{fig:timeave}}
\end{figure}

The numerical results in the adaptive circuit are obtained by direct simulation of individual trajectory dynamics resulting from applying unitary gates, random measurement processes and adaptive qubit rotations depending on the measurement outcomes. Individual measurement outcomes are obtained by drawing them from a two-state distribution determined by the Born rule probabilities of 0 and 1 states. In the final time step, all qubits are measured according to this prescription. The charge distributions are obtained by repeating the process multiple times, closely following the experimental approach introduced in the main text. The only difference between the numerical treatment and the experimental protocol is that, to mitigate the heavy calculations, we complement the repetitions of different trajectory realizations by carrying out temporal averaging. Instead of simply averaging the observables over $N$ trajectory realizations, as in an experiment, in the calculation we also average over $T$ time steps, taking a measurement at the end of each cycle. This approach should be regarded as a purely numerical trick and corresponds to simulating effectively $NT$ individual trajectories if the states at different times obey similar statistics as different repetitions of the circuit execution. The validity of this procedure is illustrated in Fig.~\ref{fig:timeave}, showing that the results obtained by averaging over with $N$ trajectories match the results obtained by averaging over $T$ time steps after the system has reached the steady state. Thus, the effective number of repetitions $NT$ reported in the main text matches closely the actual number of repetitions.

\section{Note on single-parameter scaling}\label{si:scaling} \label{app:B}

 The single-parameter scaling assumption $F(p,L)-F(p_c,L)=\tilde{F}\left[(p-p_c)L^{1/\nu}\right]$ near criticality is an asymptotic form approached in the thermodynamic limit. While it is a simple tool to extract qualitative information, for quantitatively accurate fitting of parameters and estimating their error bounds, it is necessary to include subleading scaling variables accommodating the finite-size corrections to the single-particle scaling. This process typically requires system sizes $L=10^2-10^3$ to yield reliable results, which is well beyond any numerical methods for strongly correlated systems. In particular, the size limitation $L\leq 20$ for the present problem makes it impossible to extract quantitatively accurate values for $p_c$ and $\nu$. We emphasize that the problem is not technical but fundamental—the single parameter Ansatz is simply not a precise description of small systems. This limitation of numerics is encountered when attempting to demonstrate the existence of an entanglement transition from $S_\text{vN}$ as well. Therefore, limited to small systems, one should not expect that technically highly sophisticated fitting procedures could produce a more accurate result. With these caveats, we explain below how we arrived at the estimates $p_c=0.14$ and $\nu=1.3$ employed in Sec.~\ref{subsec:postsel} and \ref{subsec:steering}.      

\begin{figure}[t!]
    \centering
    \includegraphics[width=.99\columnwidth]{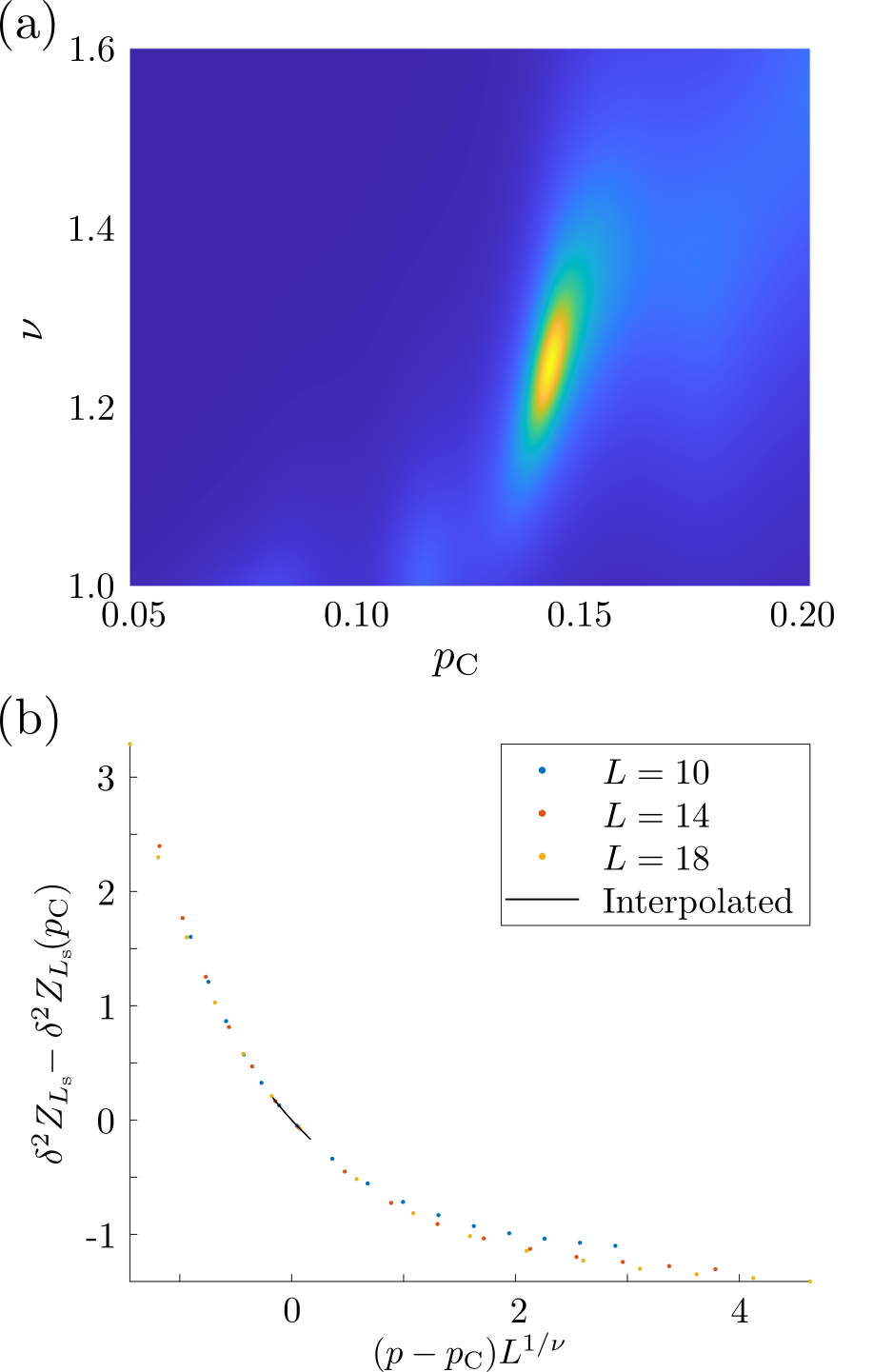}  
  \caption{  \textbf{Estimates for the critical parameters} (a) Heat plot of the inverse square sum $C^{-1}$. The yellow region corresponds to large values. (b) Scaling collapse corresponding to the maximum in (a). The solid line indicates the critical region in the fitting.  } 
    \label{fig:scaling}
\end{figure}

Here we consider the charge fluctuations illustrated in Fig.~\ref{fig:fluc1}(b). To estimate the critical parameters, we minimize the square sum $C=\sum_{x \in I}\sum_{L=10,14,18} (y(L,x) - \mu(x))^2$, where $y$ represents postselected fluctuations shifted by the $p_\text{c}$ value, and $x$ values correspond to $(p-p_\text{c})L^\frac{1}{\nu}$; $\mu(x)$ is the average $y$ over all $L$ at that point. The $x$ values have further been limited to an interval around zero corresponding to $I = [-w\text{min}_L(L^\frac{1}{\nu}),w\text{min}_{L}(L^\frac{1}{\nu})]$ with $w = 0.05$, as the fitting is only expected to work near zero and in order to ensure the interval contains data points for each $L$. Finally, the data has been interpolated within this range to obtain an equal number of terms in the sum for each point in the diagram. We have elected to use the odd subsystem lengths to obtain maximum $L$ while bypassing finite-size effects expected due to the even/odd effects. In Fig.~\ref{fig:scaling} (a) we plot the inverse of the square sum $C^{-1}$ for each pair $(p_\text{c},\nu)$, showing that the minimizing parameters are located in the vicinity of $p_c=0.14$ and $\nu=1.3$. Fig.~\ref{fig:scaling} (b) shows the scaling collapse of the points corresponding to minimizing values, obtaining an optimum at $(p_\text{C} = 0.143, \nu = 1.25)$. The critical exponent thus obtained is close to the percolation exponent $\nu = 4/3$ that would be expected on theoretical grounds \cite{li_measurement-driven_2019}. Due to system size limitations, the estimates for parameters vary $10\%-20\%$ for small variations of the critical interval or considering the even subsystem sizes instead of the odd ones. 

As discussed in Sec.~\ref{subsec:postsel}, while the single-parameter scaling for available system sizes cannot reliably determine whether the entanglement entropy and the charge fluctuations belong to the same universality class, the parameter free entropy vs. fluctuation collapse in Fig.~\ref{fig:fluc1} (c) indicates precisely that. Also, it concretely predicts that to determine the correct critical parameters via single-parameter scaling alone, one would need system sizes for which the value of charge fluctuations at $p_c$ satisfies $\delta^2Z_{L_s}\gg 2$, well beyond current capabilities.  Similarly, the accurate match between the postselected fluctuations and steered fluctuations in volume-law and area-law regimes, shown in Fig.~\ref{fig4}(c), implies that the steered fluctuations share the same critical data even though single-parameter scaling analysis cannot yield a definite conclusion.

\section{Charge fluctuations vs entanglement entropy}
\label{app:c}

\begin{figure}[ht]
    \centering
    \includegraphics[width=.99\columnwidth]{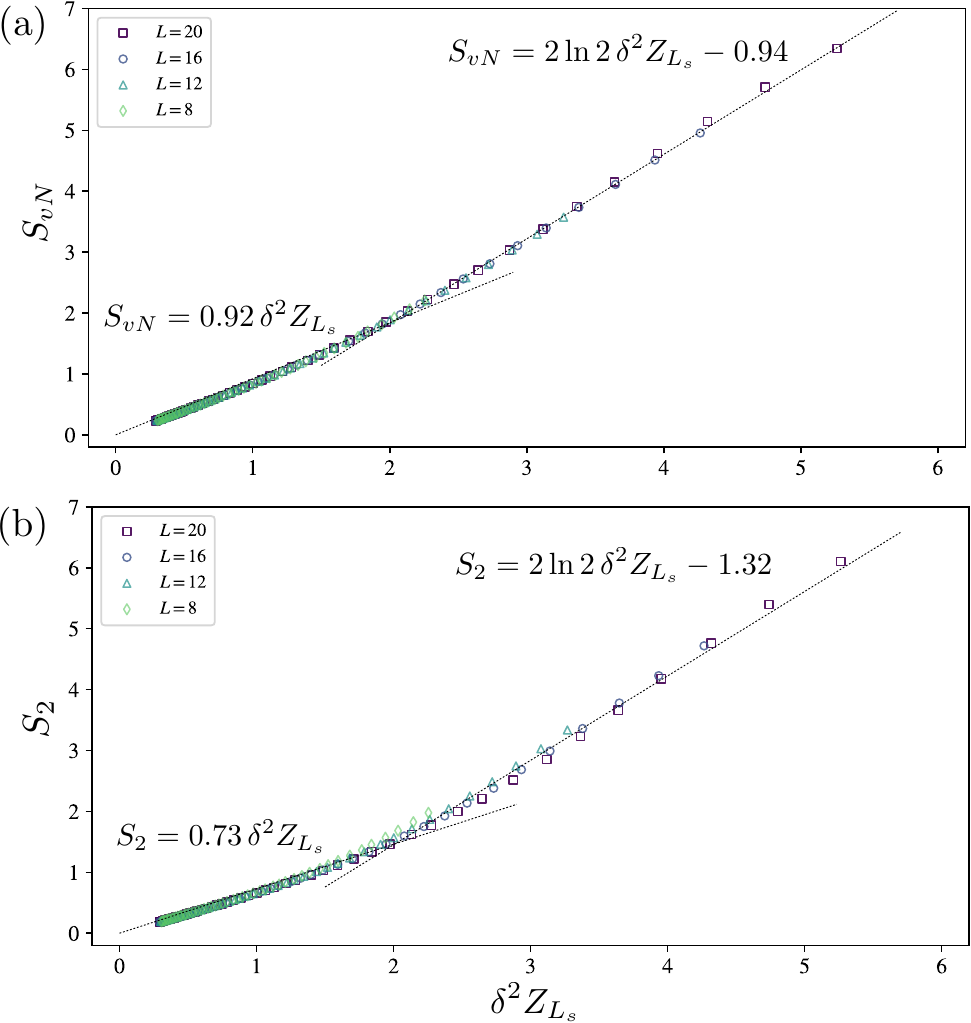}  
  \caption{  \textbf{Entanglement entropies vs. bipartite charge fluctuations} \textbf{a,} shows the von Neumann entropy as a function of bipartite charge variance (essentially the same as Fig. \ref{fig:fluc1}c). \textbf{b,} presents analogous results for the second Rényi entropy. Apart from numerical values for the slope in the low-entanglement regime and the vertical intercept in the high-entanglement regime, both von Neumann and Rényi entropies exhibit a remarkably universal dependence on bipartite charge fluctuations. } 
    \label{fig:Renyi_vN}
\end{figure}

In this section, we provide analytical results for the entanglement entropy $S_\text{vN}$ and the subsystem charge fluctuations $\delta^2 Z_{L_s}$ in $U(1)$-charge conserving random unitary circuits without measurements. As illustrated in Fig.~\ref{fig:fluc1}c in the main text, the entropy-fluctuation relation is universal and does not depend on the system size or the single-qubit measurement rate. This conclusion extends to higher R\'enyi entropies, defined as 
${\cal S}_{n}=\frac{1}{1-n}\ln\left[{\rm Tr}(\rho_s^n)\right] $ where the specific case of $n\to 1^+$ is the standard von Neumann entropy. This behavior is demonstrated in Fig.~\ref{fig:Renyi_vN} for the second R\'enyi entropy together with the von Neumann entropy. This observation has significant implications. As elaborated below, we perform analytical calculations in the $p\to 0^+$ limit, successfully deriving the asymptotic form of the entropy versus fluctuations curve analytically. This leads to the remarkable conclusion that the R\'enyi entropies exhibit the asymptotic form ${\cal S}_{n}\to 2\ln 2 \,\delta^2Z_{L_s}+\mathcal{O}(1)$ for $\delta^2 Z_{L_s}\gg 1$, where the latter term represents a subleading correction that varies with the R\'enyi index $n$. The proportionality for large values indicates that the entropies and charge fluctuations undergo a volume-area phase transition at the same critical value $p_c$, belonging to the same universality class. Consequently, the collapse of the entropy versus fluctuations data onto a single curve (without any fitting parameters) provides compelling evidence that fluctuations serve as a reliable probe of the entanglement entropy phase transitions in the system under study. In App.~\ref{app:D}, this numerical observation is supported by a theoretical argument.

 We now derive the entanglement entropy and the charge fluctuations in the $p=0$ case. This analysis essentially determines the relationship between $S_\text{vN}$ and $z_{L_s}$ in the volume-law regime. For the sake of convenience, we first derive the results for a modified form of the charge operator, $Q_L = (L + Z_L)/2$, which corresponds to labeling qubit charge eigenvalues as $\{0,1\}$ instead of $\{-1,1\}$. The same form is also used for the subsystem charge and the variances will be simply related to each other as $\delta^2 Q_{L_s}=(1/4)\delta^2 Z_{L_s}$. 

We first note that the sizes of different sectors of the Hilbert space corresponding to total charges ${\cal Q}_L \in (0, \cdots, L)$ are given by
\begin{align}
    {\cal N}_{{\cal Q}_L} = {{L}\choose{{\cal Q}_L}} = \frac{L!}{{\cal Q}_L!(L-{\cal Q}_L)!},
\end{align}
which sum to the full Hilbert space size $\sum_{{\cal Q}_L=0}^{L}{\cal N}_{{\cal Q}_L} = 2^{L}$. A generic random state in a charge sector ${\cal Q}_L$, which is a superposition of all possible basis states with uniformly distributed random coefficients, can be written as
\begin{align}
    \ket{\Psi} = \sum_{\{q_i\}}^\prime c_{q_1, \cdots, q_L} \ket{q_1, \cdots, q_L},
\end{align}
with the summation constrained over $q_i=0,1$ ($i=1, \cdots, L$) such that $\sum_i{q_i}={\cal Q}_L$. For a random state with a uniformly distributed superposition, the coefficients on average satisfy $|c_{q_1, \cdots, q_L}|^2 \sim 1/{\cal N}_{{\cal Q}_L}$. Considering the charge sector ${\cal Q}_L = L/2$ (or $Z_L=0$), and half-partitioning the system, the reduced density matrix for each subsystem is block-diagonal, consisting of $L/2$ blocks. Each block corresponds to a different charge sector ${\cal Q}_{L_s} \leq L/2$ of the subsystem with a dimension:
\begin{align}
    {\cal N}_{{\cal Q}_{L_s}} = {\frac{L}{2}\choose{{\cal Q}_{L_s}}}.
\end{align}
Within each of these blocks, the eigenvalues of the reduced density matrix are approximately equal due to the random nature of the state. Using a simple counting argument elaborated below, we estimate them as
\begin{align}
    p_{{\cal Q}_{L_s},i} = \frac{1}{{\cal N}_{{\cal Q}_L}} \binom{\frac{L}{2}}{ \frac{L}{2} - {\cal Q}_{L_s} }, \label{eq:micro-probabilities}
\end{align}
where ${\cal N}_{{\cal Q}_L} = \binom{L}{ \frac{L}{2} } $ is the total number of configurations with total charge $ {\cal Q}_L = L/2 $. The factor $ 1/{\cal N}_{{\cal Q}_L} $ arises from the normalization of the wave function coefficients. The binomial coefficient counts the number of distinct charge configurations in the complementary subsystem, which contains $L/2$ qubits. The charge of the complementary subsystem is ${\cal Q}_{\mathrm{comp}} = {\cal Q}_L - {\cal Q}_{L_s} = L/2 - {\cal Q}_{L_s}$. Substituting ${\cal Q}_{\mathrm{comp}}$ into the binomial coefficient yields the expression for $p_{{\cal Q}_{L_s},i}$ above.

Applying the rule of sum of the squares of binomial coefficients 
\begin{align}
   \sum_{i =0}^{\frac{L}{2}} {L/2\choose{i}}^2= {L\choose{L/2}},
\end{align}
we can explicitly verify that the condition
\begin{align}
{\rm Tr} (\rho_s) = 
     \sum_{{\cal Q}_{L_s} =0}^{\frac{L}{2}} \sum_{i=1}^{{\cal N}_{{\cal Q}_{L_s}}}
     \: p_{{\cal Q}_{L_s},i}
    =
    \sum_{{\cal Q}_{L_s} =0}^{\frac{L}{2}}
     {\cal N}_{{\cal Q}_{L_s}}
     \: p_{{\cal Q}_{L_s},i}
   = 1,
\end{align}
is satisfied. We can also evaluate the entanglement entropy, whose leading term reads 
\begin{align} \label{eq:entropy}
{\cal S}_{vN} =-{\rm Tr} (\rho_s \ln \rho_s) \sim
    \frac{1}{2}\ln {{L}\choose{L/2}}\sim \frac{L}{2}\ln 2,
\end{align}
where terms as ${\cal O}(1)$ or smaller are dropped assuming $L\gg1$.

To study the subsystem charge fluctuations, we consider the coarse-grained distribution for the subsystem charge which reads
\begin{align}
    P_{{\cal Q}_{L_s}} 
    =
     {\cal N}_{{\cal Q}_{L_s}}
     \: p_{{\cal Q}_{L_s},i} =
    \frac{ {L/2\choose{{\cal Q}_{L_s}}} {L/2\choose{ L/2 - {\cal Q}_{L_s}}}  }{ {L \choose L/2}}. 
\end{align}
This distribution is symmetric under ${\cal Q}_{L_s} \leftrightarrow L/2-{\cal Q}_{L_s}$ as it is also expected by simply interchanging the two subsystems with each other. 
Hence, the average charge within subsystems should also satisfy the symmetry which implies $\langle {\cal Q}_{L_s} \rangle = L / 4$. 
Through simple algebra and the rule for the sum of squares of binomial coefficients, we find that
\begin{align}
    \langle {\cal Q}_{L_s}^2  \rangle 
    &=\sum_{{\cal Q}_{L_s}=0}^{L/2} P_{{\cal Q}_{L_s}} 
    {\cal Q}_{L_s}^2  =\sum_{{\cal Q}_{L_s}=0}^{L/2} 
    \frac{ {L/2\choose{{\cal Q}_{L_s}}}^2  }{ {L \choose L/2}} {\cal Q}_{L_s}^2  
    \nonumber\\
    &=    \frac{(\frac{L}{2})^2   }{ {L \choose L/2}} 
    \sum_{{\cal Q}_{L_s}=1}^{L/2}   {L/2-1\choose{{\cal Q}_{L_s}-1}}^2
        \nonumber\\
    &= \frac{(\frac{L}{2})^2   }{ {L \choose L/2}} 
      {L-2\choose{L/2-1}} = \frac{L^2}{16}\frac{L}{L-1}
\end{align}
from which the variance of charge fluctuations are obtained
as $    \delta^2{\cal Q}_{L_s} =(1/16)L^2/(L-1)$. Consequently, we have  
$    \delta^2 Z_{L_s} =(1/4)L^2/(L-1)\sim L/4$
for original charge quantities $Z_{L_s}$ used in the main text. Finally, by employing \eqref{eq:entropy}, we obtain the fluctuation-entropy relation,
\begin{equation}
   {\cal S}_{vN}=2\ln 2 \,\delta^2Z_{L_s}+\mathcal{O}(1), 
\end{equation}
which becomes an accurate description in the regime $\delta^2Z_{L_s}>2$, as seen in Fig.~\ref{fig:Renyi_vN}a. Also, since the different R\'enyi entropies share the same maximum value $L/2\ln 2$ saturated at $p=0$ (modulo different sub-leading corrections), the result also holds for higher R\'enyi entropies, as illustrated in Fig.~\ref{fig:Renyi_vN}b for the second R\'enyi entropy. In the small fluctuation regime $\delta^2Z_{L_s}<2$, the entropy-fluctuation relations are also linear to a good approximation, but the slope does not have a universal value for different entropies.

\section{The argument for coinciding phase diagrams of the entanglement entropy and the bipartite charge fluctuations}\label{app:D}

Here, we establish a general argument that the entanglement entropy and bipartite charge fluctuations share the same phase diagram in the presence of postselection. We demonstrate this in two parts: first, we show that the entanglement area-law always implies a charge fluctuation area-law, and second, that the entanglement volume-law for a random brickwork circuit implies a charge fluctuation volume-law. Together, these two facts necessitate that the phase transitions for entanglement entropy and charge fluctuations must coincide.

\subsection{Entanglement area law implies charge fluctuation area law } \label{subsec:area_law}

Let us first demonstrate that the charge fluctuations of a subsystem $A$ in a bipartite setting can be expressed in terms of the connected correlation between the subsystem charge $Z_{A}$ and the charge of its complement $Z_{B} = Q - Z_{A}$, where $Q$ represents the fixed total charge:
\begin{align}
\delta^2 Z_{A} &= \langle Z_{A}^2 \rangle - \langle Z_{A} \rangle^2 \nonumber \\
&= \langle Z_{A} (Q - Z_{B}) \rangle - \langle Z_{A} \rangle \left( Q - \langle Z_{B} \rangle \right) \nonumber \\
&= -\langle Z_{A} Z_{B} \rangle + \langle Z_{A} \rangle \langle Z_{B} \rangle \equiv -\langle Z_{A} Z_{B} \rangle_c,
\end{align}
where $\langle AB \rangle_c = \langle AB \rangle - \langle A \rangle \langle B \rangle$ denotes the connected correlation function. We note that $\delta^2 Z_{A} \geq 0$ implies an anticorrelation between the charges of the two subsystems: $\langle Z_{A} Z_{B} \rangle_c \leq 0$.

In the area-law entanglement regime, all connected correlation functions are expected to be short-ranged:
\begin{align}\label{eq:damping1}
\langle Z_{i} Z_{j} \rangle_c \sim e^{-|i - j| / \xi}.
\end{align}
As the subsystem charge fluctuations can be rewritten in terms of such functions,
\begin{align}\label{eq:sumform1}
\delta^2 Z_{A} = -\langle Z_{A} Z_{B} \rangle_c = -\sum_{i \in A} \sum_{j \in B} \langle Z_{i} Z_{j} \rangle_c,
\end{align}
its significant contribution will only come from the sites $i$ and $j$ near the boundary of $A$ and $B$, within a layer of width $\sim\xi$. This distance is independent of subsystem size, and as such the charge fluctuations also exhibit an area-law behavior when we have area-law entanglement.

\subsection{Entanglement volume law implies charge fluctuation volume law in brickwork circuits }

The argument that the entanglement volume-law phase implies a charge fluctuation volume-law phase is more involved and hinges on the structure of the random brickwork circuit. Although it is in general possible to construct states exhibiting an entanglement volume-law and a charge fluctuation area-law, such states are highly unlikely to occur in brickwork dynamics. In fact, the circuit cannot simultaneously sustain an entanglement volume-law phase and a charge fluctuation area-law phase. To elucidate the structure of the argument, we divide it into the following steps, which will be detailed next:

\begin{enumerate}
    \item The charge fluctuation area-law in the brickwork circuit requires that the averaged (semi)-local charge correlation functions are short ranged $\overline{\langle Z_{\mathbbm i}Z_{\mathbbm j}\rangle_c }\sim e^{-|{\mathbbm i}-{\mathbbm i}|/\xi}$. 
    Here, $Z_{\mathbbm i} $ represents the \emph{slightly coarse-grained} charge corresponding to the charge of a few qubits around site $i$, and $\overline{O} = \frac{1}{\mathcal{N}_{\mathbf{U}}} \sum_{\mathbf{U}} O_{\mathbf{U}}$ denotes the average over random unitary realizations.
    \item{States that exhibit volume-law entanglement but display area-law bipartite charge fluctuations are extremely unstable under the dynamics of random local unitaries arranged in a brickwork circuit. Even a single layer of two-qubit gates can push such states into a charge volume-law phase. Moreover, returning to such an initial state during the course of time evolution is highly improbable, analogous to the Poincar\'e recurrence theorem. Overall, the chance of encountering such finely tuned states is negligible during evolution under local random unitary operations.}
    \end{enumerate}

\subsubsection{The charge area-law phase supports only short-range charge correlations}

In Sec.~\ref{subsec:area_law}, we observed how area-law (short-range) entanglement necessitates the same behavior for charge fluctuations. Conversely, we can show that area-law charge fluctuations imply short-range correlations on average. To this end, we rewrite the  total charge fluctuation as
\begin{align}
\delta^2 Z_T = \sum_{\mathbbm{i}, \mathbbm{j}} \langle Z_{\mathbbm{i}} Z_{\mathbbm{j}} \rangle_c = 
 \sum_{\mathbbm{i}} \langle Z_{\mathbbm{i}}^2 \rangle_c + \sum_{\mathbbm{i} \neq \mathbbm{j}} \langle Z_{\mathbbm{i}} Z_{\mathbbm{j}} \rangle_c = 0,
\end{align}
in terms of the slightly coarse-grained local charge variables
\begin{align}
Z_{\mathbbm{i}} = \sum_{n \in (i-\frac{w}{2}, i+\frac{w}{2})} Z_n,
\end{align}
for a small number ($w$) of qubits centered around the site $i$. 
\par
Since $\langle Z_{\mathbbm{i}}^2 \rangle_c \geq 0$, the sum of (semi)-local charge correlations corresponding to non-diagonal terms is non-positive for each trajectory and within every realization of random unitaries:
\begin{align}
\sum_{\mathbbm{i} \neq \mathbbm{j}} \langle Z_{\mathbbm{i}} Z_{\mathbbm{j}} \rangle_c \leq 0.
\end{align}
For a particular realization of random unitaries, some charge correlation terms inside the sum might be positive. Nevertheless, these positive values do not persist after averaging over different random unitaries because the two-qubit unitaries are uncorrelated and their sole generic property is the charge conservation. Therefore, all of the averaged correlations of (semi)-local charges are non-positive: $\overline{\langle Z_{\mathbbm{i}} Z_{\mathbbm{j}} \rangle_c} \leq 0$.

In a bipartite one-dimensional systems, the charge fluctuation area-law phase is characterized by the property that, the average charge fluctuations of a subsystem become independent of the subsystem size once it exceeds a characteristic length scale $\xi$:
\begin{align}
|\overline{\delta^2 Z_{A}}-\delta^2 Z_{A,\infty}|\leq e^{-L_A/\xi}  
\end{align}
where $L_A$ is the size of the (smaller) subsystem, and $\delta^2 Z_{A,\infty}$ represents the thermodynamic limit value of charge fluctuations. Similar to Eq. \eqref{eq:sumform1}, the charge fluctuations can be described in terms of correlations:
\begin{align}
\overline{\delta^2 Z_{A}} = -\sum_{{\mathbbm i} \in A, {\mathbbm j} \in B} 
\overline{\langle Z_{\mathbbm i} Z_{\mathbbm j} \rangle_c}
\end{align}
As we saw above, due to charge anti-correlations, each term in the above sum is non-positive when averaged over unitary realizations. Consequently, the charge area-law implies that these individual terms must also be short-ranged:
\begin{align}
\overline{-\langle Z_{\mathbbm i} Z_{\mathbbm j} \rangle_c} \sim e^{-|{\mathbbm i} - {\mathbbm j}|/\xi}.
\end{align}
Therefore, in random brickwork circuits, the charge area-law phase permits only short-range charge correlations.

\subsubsection{Long-range entanglement inevitably leads to long-range charge correlations}

As noted above, in principle, states with vanishing bipartite charge fluctuations may yet support potentially large entanglement. Such states can typically be written as
$
\ket{\Psi} = \sum_k c_k \ket{\psi_k^A} \otimes \ket{\psi_k^B},
$
where $ Z_A \ket{\psi_k^A} = q_A \ket{\psi_k^A} $ and $ Z_B \ket{\psi_k^B} = q_B \ket{\psi_k^B} $ for fixed subsystem charges $q_A$ and $q_B$, with potentially additional terms of different charge concentrated near the subsystem edges. 
However, we now show that, due to the structure of the random brickwork circuit, such states will be immediately dissolved into states exhibiting long-range charge correlations.

\begin{figure}[t!]
    \centering
    \includegraphics[width=0.99\columnwidth]{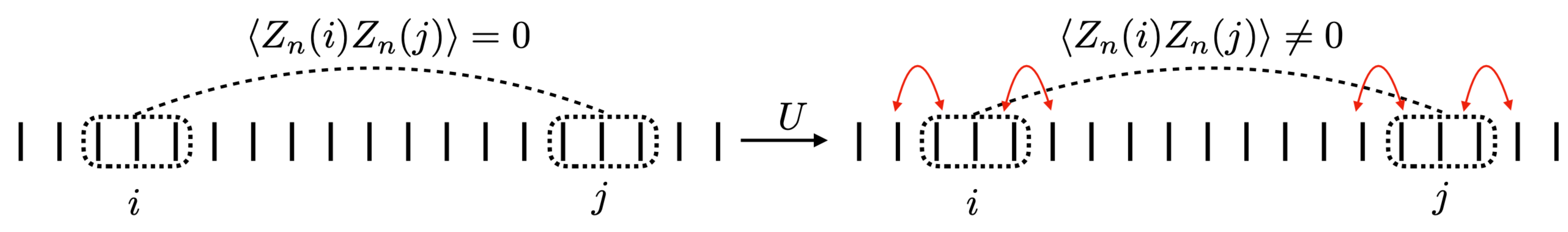}  
    \caption{  \textbf{Generation of long-range charge fluctuations from long-range entanglement.} A state which exhibits long-range entanglement but short-range charge fluctuations (left) is transformed to a state with long-range charge correlations by a single layer of brickwork circuit (right). This happens because the long-range entangled clusters with initially non-fluctuating charge states couple to the adjacent charge degrees of freedom (red arrows), switching on the long-range charge correlations. }      
    \label{fig:longrange}
\end{figure}

Suppose that at some time and for a given realization of random unitaries, the circuit has evolved into a volume-law entangled state with at most short-range charge correlations, $\langle Z_{\mathbbm{i}} Z_{\mathbbm{j}} \rangle_c \sim e^{-|{\mathbbm{i}} - {\mathbbm{j}}|/\xi}$,
as required in the charge area-law phase. The volume-law entanglement necessitates that some correlations $\langle A_{\mathbbm{i}} B_{\mathbbm{j}} \rangle_c$ do not decay exponentially. For these states, assuming clusters at positions $\mathbbm{i}$ and $\mathbbm{j}$ are sufficiently far apart, the vanishingly small charge correlations imply that at least one of the two clusters must have an almost well-defined charge ($\delta^2 Z_{\mathbbm{i}} \approx 0$ or $\delta^2 Z_{\mathbbm{j}} \approx 0$).

\begin{figure}[t!]
    \centering
    \includegraphics[width=0.99\columnwidth]{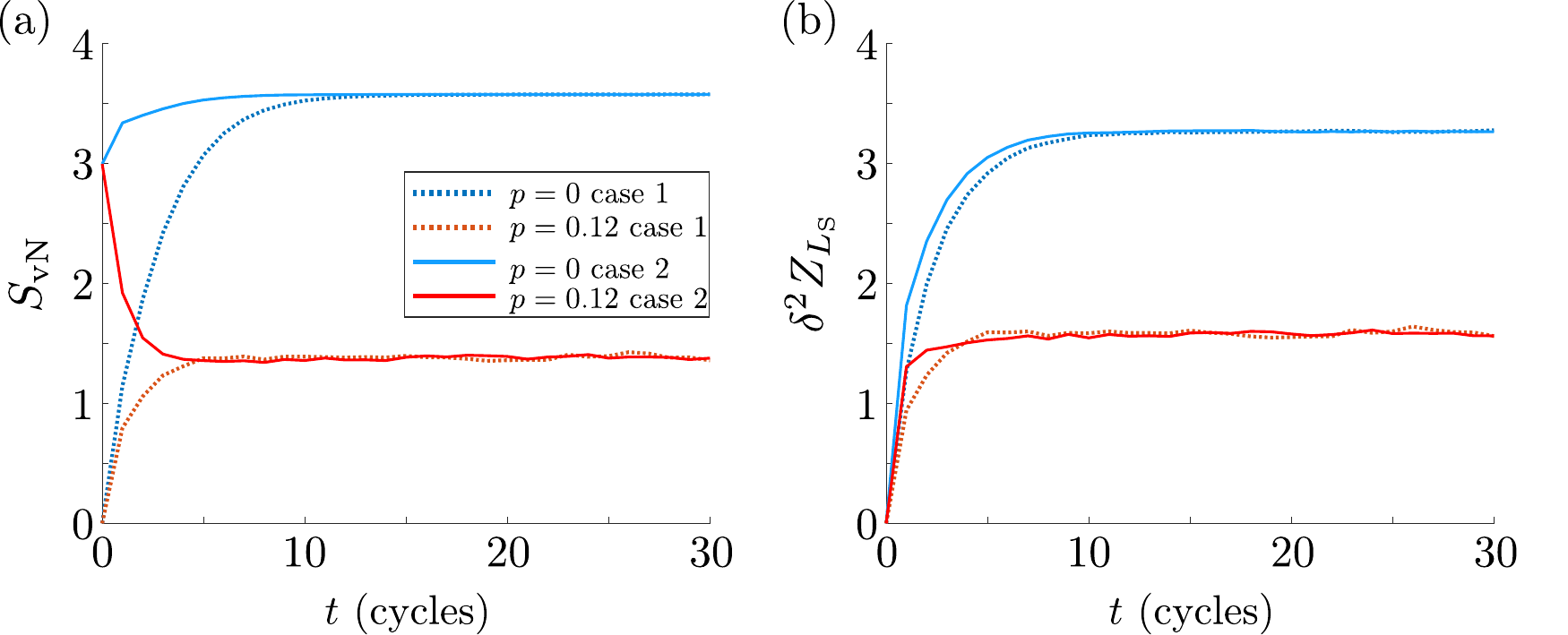}  
    \caption{ \textbf{Volume-law entanglement induces volume-law charge fluctuations.} Time evolution of the entanglement entropy (a) and the post-selected charge fluctuations (b) starting from a N\'eel state (case 1, dotted curves) and an initial state with volume-law entanglement and vanishing charge fluctuations (case 2, solid lines) at time $t=0$ for systems size $L=12$ (averaged over 1000 configurations). After a single cycle at $t=1$, the charge fluctuations have jumped to finite value which is a sizable fraction of the steady state value. The volume-law initial state in was generated through an equal superposition of all pure states with zero subsystem charges, mirrored along the subsystem boundary.}
    \label{fig:time_evo}
\end{figure}

Now, in the Heisenberg picture, applying a single layer of the brickwork dynamics transforms the local charge operators as $Z'_{\mathbbm{i},\mathbbm{j}} = {\bf U}^\dagger Z_{\mathbbm{i},\mathbbm{j}} {\bf U}$, where the unitary operator ${\bf U}$ for a single layer is a direct product of a sequence of local two-qubit unitary gates. Thus, from the perspective of the clusters at $\mathbbm{i}$ and $\mathbbm{j}$, these unitaries couple them to their neighboring qubits at the boundaries of the clusters, as illustrated in Fig.~\ref{fig:longrange}. We note that two-qubit boundary unitaries between each cluster and its neighbor necessarily exist in either odd or even half-cycles. Obviously, these local transformations cannot destroy or influence the long-range entanglement between two distant clusters. However, the boundary unitary gates generally mix different charge states of each cluster and do not conserve local charges, i.e.,
$ [{\bf U}, Z_{\mathbbm{i}}] \neq 0 \quad \text{and} \quad [{\bf U}, Z_{\mathbbm{j}}] \neq 0$.
Consequently, the long-range coupled clusters now exhibit charge fluctuations. This elucidates how any long-range correlations, after a single layer of unitaries, also induce long-range charge correlations $\langle Z'_{\mathbbm{i}} Z'_{\mathbbm{j}} \rangle_c \neq 0$ even far beyond the previous correlation length scale $\xi$. The rapid onset of charge fluctuations starting from a volume-law entangled state with initially vanishing charge fluctuations is illustrated in Fig.~\ref{fig:time_evo}(a).  

Importantly, we can see the above process is fundamentally non-reciprocal in the following sense. Once a single layer of unitaries has transformed the state from having short-range to long-range charge correlations, the probability of returning to the initial state with another layer of unitaries is negligible. This is because the second layer of random unitaries would need to exactly counteract the effects of the first layer, which coupled the clusters to their neighboring sites. The likelihood of even approximately canceling the long-range charge fluctuations in a few steps is extremely low. For all practical purposes, the onset of long-range charge correlations in a volume-entangled state is irreversible, as demonstrated in Fig.~\ref{fig:time_evo}(b), where charge fluctuations rapidly build up and reach their maximum after very few layers of unitaries. Indeed, from the figure it is clear that due to the mechanism outlined here, the presence of these long-range correlations also results in a shorter saturation time for charge fluctuations than when starting from an unentangled state. Once these charge fluctuations have saturated to their volume-law value, they remain at this level for any practically feasible timescale. 

It is insightful to note that the process of returning to a fine-tuned state with volume-law entanglement and area-law charge fluctuations is analogous to Poincar\'e recurrence for a box of gas, where molecules initially occupy one corner. The timescales associated with such recurrences are unimaginably longer than those in any conceivable experiment or simulation. Consequently, the fraction of fine-tuned configurations encountered during time evolution is extremely small compared to generic states that exhibit volume-law scaling for both entanglement entropy and charge fluctuations.

As demonstrated in App.~\ref{app_A} section and illustrated in Fig.~\ref{fig:timeave}, averaging observables over different trajectories (after the initial transient) is equivalent to averaging over time for a single trajectory. Since the frequency of visiting fine-tuned states during time evolution is negligible, they contribute insignificantly to time averages and, therefore, to averages over trajectories. Thus, we can conclude that the brickwork circuit cannot simultaneously sustain an entanglement volume-law and a charge area-law phase.

\section{Properties of the steered ensembles} \label{app:E}
\begin{figure*}[ht]
    \centering
    \includegraphics[width=1.95\columnwidth]{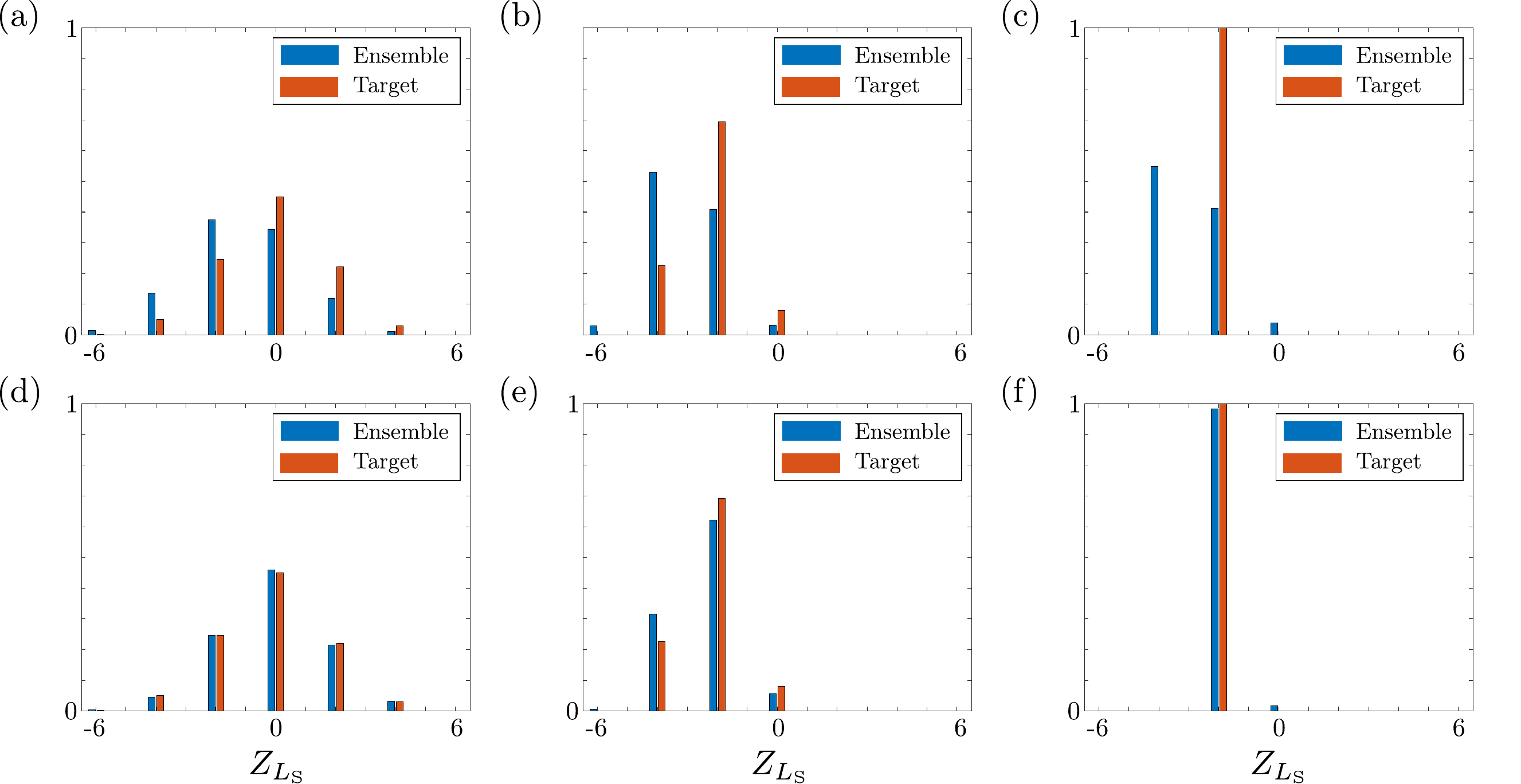}  
    \caption{\textbf{Subsystem charge distribution of the steered ensembles.} \textbf{a-c,} Comparison between the expected subsystem $S_z$ distribution of a single postselected target trajectory and the measured subsystem spins for 5000 trajectories steered towards this state. Each presents a single target state for an $L=12$ system at $p = 0.05, 0.145$ and $0.5$ respectively. \textbf{d-f,} Same as above, but using only the data from the $S_{z,\mathrm{tot}} = 0$ subset of the (same) steered trajectories.
    \label{fig:spindist_subsyst}}
\end{figure*}

Here we illustrate the key statistical properties of the steered ensembles. Since the subsystem charge distributions of a single trajectory encodes its entanglement entropy, these distributions are the main focus here.

In Fig.~\ref{fig:spindist_subsyst}(a)-(c), we present the comparison of charge distributions of a single target state $\ket{\Psi}_{{\bf m}}$ and the corresponding steered ensemble $\rho_{\bf U,m}$. A single target distribution is sensitive to the specific measurement outcomes and, in general, exhibit little symmetry. This indicates that single-trajectory distributions exhibit significant case-to-case fluctuations. The steered charge distributions corresponding to the target are clearly strongly correlated with the target distribution. This correlation is qualitatively better in the volume-law regime.

As Fig.~\ref{fig:spindist_subsyst}(d)-(f) illustrates, the correlation between the target distribution and the steered distribution is dramatically improved by considering only the steered trajectories which end in the same total charge sector $Z_L=0$ in which the target trajectory belongs to. In this case the steered distribution is obtained from density matrix $\rho_{\bf U,m}^{Z_L=0}$. Deep in the volume-law regime, the match is essentially perfect, but becomes less exact when moving towards the area-law regime.

While the distributions corresponding to individual target trajectories have little symmetry and exhibit strong case-to-case fluctuations, Fig.~\ref{fig:spindist_total} illustrates how the steered ensemble averaged over many target trajectories are smooth and symmetric.  This is just a reflection of the standard postselection problem for nonlinear quantities: it is crucial to first calculate the charge variance from a steered ensemble corresponding to a single target, and only afterwards average over steered ensembles corresponding to different target trajectories.

\section{Unraveling coherent and incoherent subsystem charge fluctuations}\label{app:F}

A general counting argument put forward in Ref.~\cite{PhysRevResearch.4.023200} suggests that the fluctuations of a conserved extensive charge in a bipartite system exhibit the same spatial scaling with the entanglement entropy. This observation connects entanglement and fluctuations in pure states. Here we explore how the charge fluctuations of a chosen  target state $\ket{\Psi}_{{\bf m}}$ can be extracted from the mixed state density matrix  $\rho_{\bf U,m}=\frac{1}{\mathcal{N}_s}\sum_i 
\ket{\Psi_i} \bra{\Psi_i} 
$, which is obtained by running the steered dynamics $\mathcal{N}_s$ times with the resulting trajectories  $\ket{\Psi_i}\equiv\ket{\Psi_{{\bf m}^\prime_i\to{\bf m}}} $. Here ${\bf m}^\prime_i$ denotes the measurement outcomes of  $\ket{\Psi_i}$  while ${\bf m}$ denotes the measurement outcomes of the chosen target. For large number of steering realizations, this density matrix approaches to 
\begin{equation}\label{eq:steer2}
\rho_{\bf U,m}=\sum_i p_i
\ket{\Psi_i} \bra{\Psi_i},
\end{equation}
where $p_i$ is the probability of trajectory $\ket{\Psi}_{i}$ in the steered ensemble. The subsystem $U(1)$ charge fluctuations become
\begin{equation}\label{eq:fluct1}
\delta^2 Z_{L_s}=\langle Z_{L_s}^2 \rangle-\langle Z_{L_s}\rangle^2,
\end{equation}
where $Z_{L_s}= \sum_{n\in L_s} Z_{n}$ and $Z_n$ is the Pauli $z$ matrix operating on the $n$th qubit. 
The first term can be written as 
\begin{equation}\label{eq:s1}
\langle Z_{L_s}^2 \rangle=\mathrm{Tr} \left(\rho_{\bf U,m} Z_{L_s}^2\right)=\mathrm{Tr} \left[\sum_i p_i \ket{\Psi_i}\bra{\Psi_i} Z_{L_s}^2\right].
\end{equation}
The trace is conveniently evaluated in the charge basis $|\vec{z}_L\rangle$, where $\vec{z}_L=(\sigma_1,\sigma_2\ldots \sigma_L)$ with $\sigma_n=\pm 1$, and for which  $Z_{L_s}|\vec{z}_L\rangle=z_{L_s}|\vec{z}_L\rangle$ with $z_{L_s}=\sum_{n\in L_s}\sigma_n$. Then Eq.~\eqref{eq:s1} becomes
\begin{equation}\label{eq:s2}
\langle Z_{L_s}^2 \rangle=\sum_{\vec{z}_L,i} p_i |\langle \vec{z}_L\ket{\Psi_i}|^2 z_{L_s}^2=\sum_{\vec{z}_L} p_{\vec{z}_L} z_{L_s}^2,
\end{equation}
where 
\begin{equation}\label{eq:s3}
p_{\vec{z}_L}=\sum_{i} p_i |\langle \vec{z}_L\ket{\Psi_i} |^2.
\end{equation}
In this compact notation, the variance can be expressed as
\begin{equation}\label{eq:fluct2}
\delta^2 Z_{L_s}=\frac{1}{2}\sum_{\vec{z}_L,\vec{z}_L'} p_{\vec{z}_L} p_{\vec{z}_L'}(z_{L_s}-z_{L_s}')^2.
\end{equation}
We can identify two distinct types of terms that contribute to charge fluctuations \eqref{eq:fluct2} that we call coherent and incoherent. Each term in \eqref{eq:fluct2} involves a product of two probabilities $p_{\vec{z}_L} p_{\vec{z}_L'}$, which themselves contain contributions from all possible final states $|\psi_i(t)\rangle$, as dictated by Eq.~\eqref{eq:s3}. The state diagonal terms $i=j$ in the product are the coherent contributions and the off-diagonal terms $i\neq j$ correspond to the incoherent contributions. In a pure state, only coherent fluctuations are present and they encode the entanglement information, as illustrated by the entropy-fluctuation correspondence with the full postselection. The incoherent fluctuations arise from the statistical mixture in $\rho_{\bf U,m}$, and may not reflect the entanglement properties of individual trajectories. In order to probe the entanglement entropy, the incoherent contributions need to be subtracted. 

\begin{figure*}[ht]
    \centering
    \includegraphics[width=1.7\columnwidth]{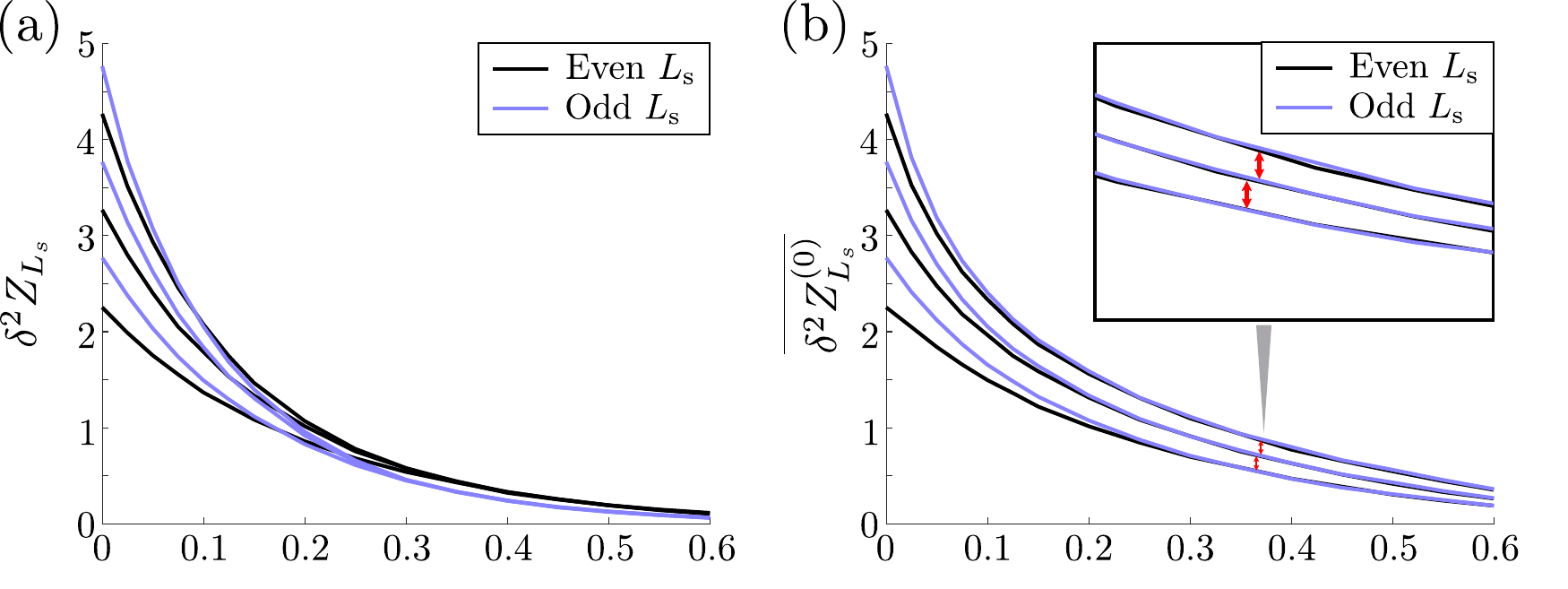}  
    \caption{\textbf{Effects of length parity on the fluctuations.} \textbf{a,} Variance as a function of $p$ for postselected data for $L = 8,12,16$ (black) and $L = 10,14,18$ (blue). In the area-law phase, the lines converge to a parity-dependent value that does not otherwise depend on length. \textbf{b,} Variance as a function of $p$ for the steered data with $Z_{L_\mathrm{s}} = 0$, for the same lengths. The red arrows indicate $2c_V(p)$ from Eq.~\ref{eq:cvp_general}.\label{Fig:SM_even_odd}}
\end{figure*}

Let's consider target trajectories which lie in the total charge sector $Z_L=0$. As discussed in Sec.~\ref{subsec:postsel} above, the steering processes break the $U(1)$ charge conservation and the state  have a finite weight at $Z_L=\pm 2$, $Z_L=\pm 4$...$Z_L=\pm L$. The density matrix is block diagonal in the $U(1)$ charge sectors $[\rho_{\bf U,m}, Z_L]=0$, so most of the incoherent processes can be removed by considering only the $Z_L=0$ block $ \rho_{\bf U,m}^{0}$.  This leads to charge fluctuations 
\begin{equation}\label{eq:S=0}
\delta^2 Z_{L_s}^{(0)}=\frac{1}{2}\sum_{\vec{z}_L,\vec{z}_L'\in\{Z_L=0\}} \tilde{p}_{\vec{z}_L} \tilde{p}_{\vec{z}_L'}(z_{L_s}-z_{L_s}')^2.
\end{equation}
with normalized probabilities $\tilde{p}_{\vec{z}_L}=p_{\vec{z}_L}/(\sum_{{\vec{z}_L}\in \{Z_L=0\}}p_{\vec{z}_L})$.  The projected fluctuations still contains residual incoherent contributions from distinct steering trajectories that end up in the $Z_L=0$ sector. As seen in Fig.~\ref{fig:spindist_subsyst}, the projected fluctuations $\delta^2 Z_{L_s}^{(0)}$ already provide an excellent match with the postselected fluctuations deep in the volume-law regime. However, the incoherent contributions from the steering processes, number of which scale as the system size, can be expected to give rise to an addition volume-law fluctuations in the area-law phase. Indeed, the target-trajectory averaged fluctuations 
\begin{equation}\label{eq:average}
\overline{\delta^2 Z_{L_s}^{(0)}}=\frac{1}{N}\sum_{j} \delta^2 Z_{L_s}^{(0)}(j),
\end{equation}
where the sum is over $N$ target trajectories $\ket{\Psi}_{{\bf m}_j}$, clearly display additional size-dependent fluctuations even when the target trajectories exhibit area-law fluctuations, as seen in Fig.~4(b).

Fortunately, the coherent and incoherent fluctuations can be easily distinguished by their different system-size dependence. 
\begin{equation}\label{eq:area}
\overline{\delta^2 Z_{L_s}^{(0)}}\sim c_V(p)L_{s}+c_A(p),
\end{equation}
where $c_A(p)$ is the area-law coefficient from coherent contribution and $c_V(p)$ is the parasitic volume-law coefficient from the incoherent contribution. By comparing the fluctuations of different systems sizes, one can straightforwardly obtain $c_V(p)$ and $c_A(p)$. 
As seen in Fig. \ref{Fig:SM_even_odd} (b), for any two different subsystem lengths, both being either even or odd, the change in the parasitic term is $\Delta\left( \overline{\delta^2 Z^{(0)}_{L_s}}\right) = c_V(p) \Delta L_s$. This reflects the volume-law nature of the parasitic term. Fig. \ref{Fig:SM_even_odd} (b) also indicates that there is a small offset between the trends for even and odd $L_s$, which arises because the last layer of two-qubit unitaries treat the even and odd subsystems differently. This odd-even effect is visible already in the postselected fluctuations seen in Fig. \ref{Fig:SM_even_odd}(a). The even-odd effect elucidates the expression for $c_V(p)$ introduced in the main text in Eq.~\eqref{eq:cvp} for successive lengths. Due to the volume-law character of the parasitic term other combinations consistent with the even-odd effect, of the general form
\begin{equation}
    c_V(p)= \left(\overline{\delta^2 Z_{L_s=2k}^{(0)}}-\overline{\delta^2 Z_{L_s=2k-j}^{(0)}}\right)/(j+1)\label{eq:cvp_general}
\end{equation}
with $j$ an odd integer will also work.

Finally, we note that in this charge-conserved case steering will not generate additional fluctuations at system sizes with $L_s=1,2$. While the difference in parasitic contributions between adjacent lengths is accurately described above, this causes an additional correction term relevant for small system sizes; taking it into account, the area-law regime parasitic contribution term is expressed as $c_V(p)  (L_s - 2)$. The introduction of the small offset $-2c_V(p)\sim 0.1$, while having little practical significance -- especially for system sizes $L_s \gg 2$ -- provides an excellent match between the reduced fluctuations  $\delta^2 \widetilde{Z}_{L_s}^{(0)}=\overline{\delta^2 Z_{L_s}^{(0)}}-c_V(p)\left[L_s-2\right]$ and the postselected fluctuation in the area-law regime, as seen in Fig.~\ref{fig4} (c).

\section{Estimating the steering overhead}\label{app:g}

To measure expectation values of observables, it is necessary to prepare multiple copies of a given trajectory $\ket{\Psi}_{{\bf m}}$ with measurement outcomes ${\bf m}$.  In practice, this necessitates running the circuit a number of times which scales exponentially in the system size, a fact which gives rise to the exponentially complex postselection bottleneck in the experimental studies of monitored dynamics. By the virtue of the steering approach discussed in the present work, we can circumvent this bottleneck, as well as the exponential bottleneck associated with the entanglement entropy measurement through state tomography. Crucially, these combined exponential complexities are replaced by a scalable polynomial overhead which we now establish. 

For a given target trajectory, the fluctuations are obtained from the steered ensemble by running the steering dynamics $\mathcal{N}_{s}$ times and 
measuring  all the qubits in the charge basis. If the total charge of the full circuit belongs to the sector
$Z_{L} = 0$, we also consider the measured value for the subsystem charge $Z_{L_s}=Z_i$. Then, using the values attained for $Z_i$ for the \emph{successful} runs which end up in the total charge $Z_L=0$ sector, we can estimate the charge fluctuations of subsystem using the so-called \emph{sample variance} as
\begin{align}
    \delta^2 Z_{L_s}^{(0)} = \frac{1}{2 n^2}\sum_{i,j}(Z_i-Z_j)^2 = \frac{1}{n}\sum_i \left( Z_i - \langle Z\rangle \right)^2,
\end{align}
with $\langle Z\rangle$ denoting the sample average of $Z_i$ \footnote{We note that slightly more precise expression for the sample variance is given by $\delta^2_{\rm unbiased} Z =  \delta^2 Z n/(n-1)$
as in a sample of size $n$, by calculating $\bar{Z}$ we are left with $n-1$ independent degrees of freedom.}. We see that there are basically two factors to be taken into account to estimate of number of sufficient steering runs:
\begin{itemize}
    \item 
    the fraction of successful runs (which end up in the $Z_L=0$ sector) in all runs $\mathcal{N}_{Z_L=0}/\mathcal{N}_{s}$ 
    \item
    the  sample-to-sample variations of obtained variances $\delta^2 Z_{L_s}^{(0)}$ with the sample size $\mathcal{N}_{Z_L=0}$. 
\end{itemize}

\begin{figure*}[ht]
    \centering
    \includegraphics[width=1.95\columnwidth]{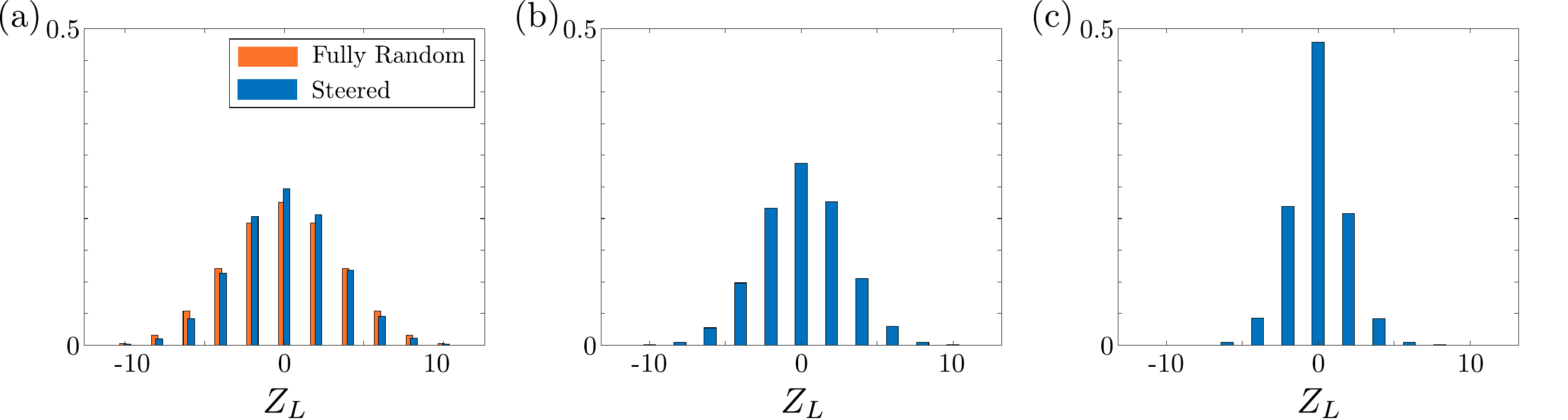}  
    \caption{\textbf{Charge distribution of the steered ensemble.} Stationary probability distributions of the total $U(1)$ charge $Z_L$ from over 5000 target trajectories with 5000 steered trajectories each for $L=12$ at probabilities \textbf{a}: $p = 0.05$, \textbf{b}: $p = 0.145$ and \textbf{c}: $p = 0.5$ respectively. For comparison, Fig.~\textbf{a} also illustrates the fully random distribution where all microstates have equal probability. 
    \label{fig:spindist_total}}
\end{figure*}

To estimate the fraction of successful runs, let us assume 
that each single qubit measurement event has almost equal probabilities for $0$ and $1$. 
By running a circuit of length $L$ and depth $T\sim L$ (to reach the steady state), at a measurement rate $p$, on average $m \sim p L T \sim p L^2$ qubits are being measured throughout the evolution.
Then, total number of different possible steering runs we can have will be ${ n }_{\rm s} = 2^m $, while only
\begin{align}
   { n}_{Z_L =0} =  {m \choose \frac{m}{2}}
\end{align}
of them are constraint to have total charge $Z_L = 0$. Therefore, the success rate of runs is given by
\begin{align}\nonumber
     \frac{{\cal N}_{Z_L=0}}{{\cal N}_{s}}=\frac{n_{Z_L=0}}{n_{s}} \sim \sqrt{\frac{2}{\pi m}} \sim \sqrt{\frac{2}{p}}\frac{1}{L}.
\end{align}
This is a rather crude approximation based on trajectories for which all measurement outcomes are equally likely, and does not account for steering which drive the trajectories towards the $Z_L=0$ sector. As the number of different charge sectors scale as $\sim L$, the $1/L$ scaling is clearly the worst case scenario. In fact, the trajectory-averaged probability distributions in Fig.~\ref{fig:spindist_total} have Gaussian-type envelope with standard deviation $\sim\sqrt{L}$. This conclusion follows from the comparison to a fully random distribution, where all microstates are equally probable. Due to the enhanced number of states in the low-lying charge sectors of the multiqubit Hilbert space, the random distribution has a standard deviation $\sim\sqrt{L}$. As depicted in Fig.~\ref{fig:spindist_total}(a), the trajectory-averaged steered distribution approaches the fully random case when $p\to 0^+$; however, the distribution remains more narrow for all $p$.  This indicates that the target-averaged probability of the $Z_L=0$ sector scales as $p_{Z=0}\sim \frac{1}{\sqrt{L}}$, hence the mean success fraction per target trajectory also scales as
\begin{align}\label{eq:success}
     \frac{{\cal N}_{Z_L=0}}{{\cal N}_{s}} \sim \frac{1}{\sqrt{L}}.
\end{align}

Next, we estimate the variance of charge fluctuations themselves by considering different samples of fixed size $\mathcal{N}_{Z_L=0}$. Ideally, we would like to reduce this ``variance of variance" below a small threshold value $\epsilon$. Intuitively, one expects that by increasing the sample size $\mathcal{N}_{Z_L=0}$ from which we calculate the subsystem charge fluctuations $\delta^2 Z_{L_s}^{(0)}$, we will have a better estimate with smaller and smaller ${\rm \delta^2} \left(\delta^2 Z_{L_s}^{(0)} \right)$. We now show that this intuition is correct and  $\delta^2 Z_{L_s}^{(0)}$ decreases with the sample size as $1/\mathcal{N}_{Z_L=0}$. In order to determine the variance of variance, we consider sets of different successful steering runs, each consisting of $n=\mathcal{N}_{Z_L=0}$ realizations which we label with capital letters to not be mixed with separate steering realizations within each set. Since every set may give a separate charge fluctuations $\delta^2_I Z$ (from here on, we drop the indices from $\delta^2 Z_{L_s}^{(0)}$ for simplicity), these charge fluctuations for different sets
basically introduce a new random variable:
\begin{align}
   Y_I = \frac{n \, \delta^2_I Z}{\delta^2_\infty Z} ,
\end{align}
where  $\delta^2_\infty Z$ is the \emph{asymptotic} value for the charge fluctuations, obtained from a infinitely large set of steering trajectories which end up in the $Z_L=0$ charge sector. The variance of variance for subsystem charge is then related to the variance of variable $Y$ (which exhibits sample-to-sample variations  between samples with  $n=\mathcal{N}_{Z_L=0}$ steering realization):
\begin{align}
    \delta^2 \left(\delta^2 Z\right) = 
    \delta^2 \left( \frac{\delta^2_\infty Z}{n}\,Y  \right) = \left( \frac{\delta^2_\infty Z}{n}  \right)^2 \delta^2 Y.
\end{align}
We can separate the contribution to this new random variable as
\begin{align}
   Y_I  &= \frac{1}{\delta^2_\infty Z}
     \sum_{i \in I}\left( Z_i - \langle Z\rangle_{I} \right)^2
     \nonumber\\
     &=
     \frac{1}{\delta^2_\infty Z}\sum_{i \in I}\left( Z_i - \langle Z\rangle_{\infty} \right)^2 -
     \frac{\left( \langle Z\rangle_{I} - \langle Z\rangle_{\infty} \right)^2}{\delta^2_\infty Z/n}\, 
     ,     \label{eq:Y_I_separated}
\end{align}
where we have used the following identity 
\begin{align}
 &   \sum_{i \in I}\left( Z_i - \langle Z\rangle_{\infty} \right)^2 = 
    \sum_{i \in I}\left( Z_i - \langle Z\rangle_{I} +\langle Z\rangle_{I}-\langle Z\rangle_{\infty} \right)^2 \nonumber\\
    & \qquad =  \sum_{i \in I}\left( Z_i - \langle Z\rangle_{I} \right)^2+ n \, \left( \langle Z\rangle_{I} - \langle Z\rangle_{\infty} \right)^2.
\end{align}

Now, the first term in the R.H.S. of Eq. \eqref{eq:Y_I_separated} can be written as the following sum over separate random variables $\sum_i^n y_i= \sum_i^n x_i^2$
where we have substituted $(Z_i - \langle Z\rangle_{\infty})/ \sqrt{\delta^2_\infty Z}= x_i$.
Assuming $Z_i$ being derived from a Gaussian statistics, then $x_i$ have a simple Gaussian distribution form
\begin{align}
    p(x) = \frac{1}{\sqrt{2\pi} }e^{-x^2/2},
\end{align}
as we have already subtracted the mean and then scaled with the standard deviation.
Using the central limit theorem, we can see that the same holds for the second term in the R.H.S. of Eq. \eqref{eq:Y_I_separated} meaning that $x_{n+1}= (\langle Z\rangle_I - \langle Z\rangle_{\infty})/ \sqrt{\delta^2_\infty Z/n}$ has a Gaussian distribution as well. As a result, the statistics of $Y_{I}$ follows from that of a random variable 
\begin{align}
    Y = \sum_i^{n} y_i - y_{n+1} = \sum_i^{n} x_i^2 - x_{n+1}^2. 
\end{align}
The lemma 1 introduced below shows that each $y_i$ above follows a $\chi^2$-distribution whose variances are $\delta^2y_i=2$.
Now treating different $y_i$'s as independent, using Lemma 2 the variance of $Y$ is found to be $2(n-1)$.

Finally, by restoring $n=\mathcal{N}_{Z_L=0}$, we find the variance of charge fluctuations is
\begin{align}
    \delta^2 \left(\delta^2 Z\right) \sim \frac{2}{\mathcal{N}_{Z_L=0}}\,\left( \delta^2_\infty Z \right)^2 ,
    \label{eq:var_of_var}
\end{align}
which requires $\mathcal{N}_{Z_L=0} > \left( \delta^2_\infty Z  \right)^2 (2/\epsilon)$ in order to have $\delta^2 \left(\delta^2 Z\right) < \epsilon$. 
This gives a lower limit to the number of successful steering runs required to 
estimate the variance up to an error $\epsilon$. Using the approximate value for the success fraction from Eq.~\eqref{eq:success}, we would need 
\begin{align}
    \mathcal{N}_{s}  \gtrsim \frac{\sqrt{L}}{\epsilon} \left( \delta^2_\infty Z  \right)^2 \sim \frac{L^{5/2}}{\epsilon}
\end{align}
which follows from the fact that the variance of the steered ensemble exhibits the volume-law scaling $\delta^2_\infty Z \propto L$.

\subsection{Lemma 1}\label{lemma1}
The square of a random variable with normalized 
     Gaussian distribution
\begin{align}
    p(x) = \frac{1}{\sqrt{2\pi} }e^{-x^2/2}.
\end{align}
     is given by $\chi^2$-distribution with one degree of freedom.
    \par
    \emph{Proof}.---Since $y=x^2$, we can have two different branches for $x$ in terms of $y$ which are simply $x_1=\sqrt{y}$ and $x_2=-\sqrt{y}$ from which we can calculate the probability distribution of $y$ as
    \begin{align}
        \tilde{p}(y) = \sum_{i} \left[ p(x) |\frac{dx}{dy}|\right]_{x=x_i(y)} =
         \frac{  p(\sqrt{y})}{\sqrt{ y}}= \frac{ e^{-y/2}}{\sqrt{2\pi y}},
    \end{align}    
which completes the proof. The cumulant generating function for $y=x^2$ can be also calculated as below
    \begin{align}
        {\cal K}_y(\theta) &= \ln \langle e^{\theta y} \rangle = \ln
        \int dy\, p(y)  e^{\theta y}  
        \nonumber\\
        &= \ln
        \int dy 
        \frac{1}{\sqrt{2\pi\, y}} e^{-y/2}  e^{\theta y} 
         = \ln\left[(1-2 \theta)^{-1/2}\right],
    \end{align}
from which we find the mean and variance to be $\langle y \rangle =1$, and $\delta^2 y = 2$, respectively.

\subsection{Lemma 2}
The cumulants of sum of independent random variables are given by the sum of the corresponding cumulants for each of those random variables. A special yet interesting case is obviously for the second cumulant as it gives the variance.  
    \par
\emph{Proof}.---
We show a similar statement is valid for the cumulant generating function and then derive it for the cumulants as they are given by $n$-th derivatives of the cumulant generating function.
The cumulant generating function for a random variable $y$ is defined as
    \begin{align}
        {\cal K}_y(\theta) = \ln \langle e^{\theta y} \rangle 
    \end{align}
If $y$ is sum of independent variables $y_i$ as $y=\sum_i y_i$ then the cumulant generating function for $y$ reads,
\begin{align}
        {\cal K}_y(\theta)   
         &= \ln \langle e^{\theta \sum_i y_i} \rangle 
         = \ln \langle \prod_i e^{\theta y_i} \rangle 
         = \ln \left(  \prod_i \langle e^{\theta y_i} \rangle \right)   \nonumber \\
         &= \sum_i \ln \langle e^{\theta y_i} \rangle = \sum_i   {\cal K}_{y_i}(\theta),
    \end{align}
which proves the statement. Note that in the last step of first line we could write the average of product of functions as the product of separate averages, since they are functions of independent random variables.

\bibliography{refs.bib}

\end{document}